\documentclass[12pt]{article}
\usepackage{amssymb,graphicx}
\usepackage{epstopdf}
\usepackage{amsmath,amsfonts}
\usepackage{epsfig}

\def\d{\partial}
\def\l{\left(}
\def\r{\right)}

\newcommand{\be}{\begin{equation}}
\newcommand{\ee}{\end{equation}}
\newcommand{\bea}{\begin{eqnarray}}
\newcommand{\eea}{\end{eqnarray}}
\newcommand{\bg}{\begin{gather}}
\newcommand{\eg}{\end{gather}}
\newcommand{\bseq}{\begin{subequations}}
\newcommand{\eseq}{\end{subequations}}

\begin{document}
\begin{flushright}
INR-TH-2011-01\\
ULB-TH/11-04
\end{flushright}
\vspace{10pt}
\begin{center}
{\LARGE \bf Scalar perturbations \\in conformal rolling scenario \\[0.3cm]
with intermediate
stage } \\
\vspace{20pt}
M.~Libanov$^{a,b}$,  S.~Ramazanov$^{a,b}$, V.~Rubakov$^{a,b}$\\
\vspace{15pt}
$^a$\textit{
Institute for Nuclear Research of
         the Russian Academy of Sciences,\\  60th October Anniversary
  Prospect, 7a, 117312 Moscow, Russia}\\
\vspace{5pt}
$^b$\textit{
Physics Department, Moscow State University,\\ Vorobjevy Gory,
119991, Moscow, Russia
}
    \end{center}
    \vspace{5pt}

\begin{abstract}

Scalar cosmological perturbations with nearly flat power spectrum may
originate from perturbations of the phase of a scalar field conformally
coupled to gravity and rolling down negative quartic potential. We
consider a version of this scenario whose specific property  is a long
intermediate stage between the end of conformal rolling and horizon exit
of the phase perturbations. Such a stage is natural, e.g., in cosmologies
with ekpyrosis or genesis. Its existence results in small negative scalar
tilt, statistical anisotropy of all even multipoles starting from
quardupole of general structure (in contrast to the usually discussed
single quadrupole of special type) and non-Gaussianity of a peculiar form.

\end{abstract}
\section{Introduction and summary}
\label{intro}

By far the most developed  hypothesis on the origin of the cosmological
perturbations is the slow roll inflation~\cite{inflation}. The
inflationary mechanism~\cite{infl-perturbations} generates almost Gaussian
scalar perturbations whose power spectrum is almost flat due to the slow
evolution of relevant parameters (the Hubble parameter and time derivative
of the inflaton field). Similar situation occurs in the inflationary
scenario with the curvaton mechanism~\cite{Linde:1996gt}; in either case,
the approximate flatness of the spectrum is a direct consequence of the
approximate de~Sitter symmetry of the inflating background.

In quest for an alternative symmetry behind the flat scalar spectrum one
naturally turns to conformal
invariance~\cite{vrscalinv,Creminelli:2010ba}. Conformal symmetry implies
scale invariance, which in the end may be responsible for the
scale-invariant scalar spectrum. An assumption of conformal invariance at
the time the primordial perturbations are generated is in line with the
viewpoint that the underlying theory of Nature may have conformal phase,
and that the Universe may have started off from, or passed through that
phase.

At the present, exploratory stage it makes sense to consider this
possibility in the context of toy models. One such model is proposed in
Ref.~\cite{vrscalinv}. Besides conventional Einstein gravity and some
matter that dominates the cosmological evolution, its main ingredient is a
complex scalar field $\phi$ conformally coupled to gravity. Conformal
invariance implies that the scalar potential is quartic, while the
dynamics is non-trivial if its sign is negative, \be V(\phi) = - h^2
|\phi|^4 \; ,
\label{jan12-1}
\ee
where $h$ is a small parameter. One assumes that the background space-time
is homogeneous, isotropic and spatially flat,
\be
ds^2 = a^2(\eta)(d\eta^2 - d{\bf x}^2) \; .
\label{jan12-2}
\ee
Then in terms of the field
\[
\chi(\eta , {\bf x}) = a(\eta) \phi (\eta , {\bf x})
\]
the dynamics is the same as in flat space-time.
One further assumes that the classical background field $\chi_c$ is
homogeneous. As it rolls down its potential $V(\chi) = - h^2 |\chi|^4$, it
approaches the late time attractor
\be
\chi_c (\eta) = \frac{1}{h (\eta_*^{(0)} - \eta)} \; ,
\label{jan5-1}
\ee
where $\eta_*^{(0)}$ is an arbitrary real parameter (``end of roll'';
the reason for the superscript $(0)$ in notation will become clear later),
and we take $\chi_c$ real without loss of generality.

The point of Ref.~\cite{vrscalinv} is that the behavior of the
phase\footnote{The normalization here is chosen for future convenience.}
 $\theta = \sqrt{2} \; \mbox{Arg}\;\phi$ in the background \eqref{jan5-1}
is very similar to what happens at inflation to the fluctuations of a
massless scalar field minimally coupled to gravity (e.g., inflaton
itself). The phase perturbations $\delta \theta$ start off as vacuum
 fluctuations and eventually freeze  out. To the leading order in $h$, the
resulting phase perturbations are Gaussian and have flat power spectrum
\be
{\cal P}_{\delta \theta} = \frac{h^2}{(2\pi)^2} \; .
\label{jan5-2}
\ee
The latter property is a consequence of conformal invariance and
$U(1)$-symmetry $\phi \to \mbox{e}^{i\alpha} \phi$ inherent in the model.
The phase perturbations are the source of the adiabatic perturbations in
this scenario, which proceeds as follows. At large field values, the
potential $V(|\phi|)$ is assumed to be different from (\ref{jan12-1}) and
to have a minimum at $|\phi |=f_{0}$; we assume that $f_{0}\ll
M_{\mathrm{PL}}$ (see also the discussion in Section
\ref{Subsec/Pg10/1:arxiv1205/Momentum scales}), so that the contribution
of the field $\phi $ to the effective Planck mass is always negligible.
At $|\phi |\sim f_{0}$, conformal symmetry is broken, the radial field
$|\phi |$ interacts with other fields, and its oscillations about the
minimum get damped quickly enough. To be on the safe side, we assume that
the field $\phi $ is a spectator at this and earlier stages, i.e., its
energy density $\rho _{\phi }$ is small compared to the energy density
$\rho _{\mathrm{tot}}$ of matter that dominates the cosmological
evolution. This is the case provided that
\begin{equation}
|\rho _{\phi }|\sim h^{2}f_{0}^{4}\ll \rho _{\mathrm{tot}}=\frac{3}{8\pi
}M_{\mathrm{PL}}^{2}H^{2}\;.
\label{Eq/Pg3/1:arxiv1205}
\end{equation}
Then the decay products of the field $|\phi |$ do not affect the evolution
of the Universe and, furthermore, the perturbations
of $|\phi|$, that exist before the end of
rolling and disappear after $|\phi|$ gets relaxed to the minimum of
$V(|\phi|)$, do not produce substantial density perturbations in the
Universe.

Once the radial field $|\phi |$ settles down to $f_{0}$, what remains are
the perturbations of the phase, which at this point are isocurvature
perturbations. They get reprocessed into adiabatic perturbations at much
later epoch by one or another mechanism. As an example, the phase $\theta$
may be pseudo-Nambu--Goldstone field, and may serve as
curvation~\cite{Linde:1996gt,Dimopoulos:2003az}. Alternatively,
perturbations $\delta \theta$ may be converted into adiabatic
perturbations by the modulated decay
mechanism~\cite{Dvali:2003em,Dvali:2003ar}. In either case, the adiabatic
 perturbations inherit the correlation properties from the phase
perturbations (with possible additional non-Gaussianity generated at the
conversion epoch), while the amplitude of  the adiabatic perturbations is,
generally speaking, smaller than that of the phase perturbations. In view
of the latter property, we treat our only parameter, the coupling constant
$h$, as free (but small).

The scenario cannot work at the conventional hot cosmological epoch, for
the following reason. The vacuum state of the phase perturbations $\delta
\theta$ is well defined at early times provided that these perturbations
evolve in the WKB regime, which implies
\be
k(\eta_*^{(0)} - \eta) \gg 1 \; , \;\;\;\;\; \mbox{early~times} \; ,
\label{jan12-3}
\ee
where $k$ is conformal momentum. On the other hand, the property that
these perturbations are frozen out at late times holds if
\be
k(\eta_*^{(0)} - \eta) \ll 1 \;  , \;\;\;\;\; \mbox{late~times} \; .
\label{jan12-4}
\ee
So, the scenario requires that both of these inequalities are
satisfied at conformal rolling stage. This can only happen if the duration
of that stage in conformal time is greater than $k^{-1}$. For conformal
momenta of cosmological significance this means that conformal rolling
lasts longer (in conformal time) than the entire hot stage until the
present epoch. Thus, the mechanism can only work   at some pre-hot epoch
at which the horizon problem is solved, at least formally. This is similar
to most other mechanisms of the generation of cosmological perturbations
(see, however, Ref.~\cite{Mukohyama:2009gg}).

At the conformal rolling stage, the dynamics of the phase perturbations
$\delta \theta$ is governed solely by their interaction with the
background field \eqref{jan5-1} (as well as with the radial perturbations
$\delta |\chi|$, see below); the evolution of the scale factor $a(\eta)$
is irrelevant. After the end of conformal rolling, the situation is
reversed. Once the radial field $|\phi|$ has relaxed to the minimum of the
scalar potential, the phase $\theta$ is a massless scalar field minimally
coupled to gravity (this is true for any Nambu--Goldstone
field~\cite{voloshin}). Since we are talking about a yet unknown pre-hot
epoch, it is legitimate to ask what happens to the perturbations of the
phase right after the end of conformal rolling. Barring fine tuning, there
are two possibilities for the perturbations $\delta \theta$:

(i) they are already superhorizon in the conventional sense at that time,
or

(ii) they are still subhorizon.

The version (i) of the scenario has been considered in
Refs.~\cite{Libanov:2010nk,YKIS}; in that case, the phase perturbations do
not evolve after the end of the conformal rolling stage, and the
properties of the adiabatic perturbations are determined entirely by the
dynamics at conformal rolling (modulo possible non-Gaussianity generated
at the conversion epoch; the latter is not specific to the conformal
rolling scenario). To subleading orders in $h$, this dynamics is fairly
non-trivial, and the resulting effects include certain types of
statistical anisotropy~\cite{Libanov:2010nk} and
non-Gaussianity~\cite{YKIS}.

In this paper we consider the second possibility, i.e., assume that there
is a long enough period of time after the end of conformal rolling, at
which the phase perturbations remain subhorizon in the conventional sense.
Their behavior between the end of conformal rolling and horizon exit
depends strongly on the evolution of the scale factor at this intermediate
stage. In order that the flat power spectrum \eqref{jan5-2} be not grossly
modified at this epoch, the scale factor should evolve in such a way that
the dynamics of $\delta \theta$ is effectively nearly Minkowskian.
Although this requirement sounds prohibitively restrictive, there are at
least two cosmological scenarios in which it is obeyed. One is the
bouncing Universe, with matter at the contracting stage having super-stiff
equation of state, $p \gg \rho$. It is worth noting in this regard that
stiff equation of state is preferred at the contracting stage for other
reasons~\cite{smooth,ekpyro} and is inherent, e.g., in a scalar field
theory with negative exponential potential, like in the ekpyrotic
model~\cite{ekpyro-i}. It is known~\cite{ekpyro-pert-old} that in models
with super-stiff matter at contracting stage, the resulting power spectrum
of scalar perturbations is almost the same as that of massless scalar
field in Minkowski space, ${\cal P}(k) \propto k^2$. This implies that the
dynamics of the scalar field perturbations is almost Minkowskian in these
models. We discuss this point further in Appendix~A. In tractable bouncing
models like those of Refs.~\cite{starting,minus-bis,minus-bis2}, our phase
perturbations exit the horizon at the contracting stage, pass through the
bounce unaffected (cf. Ref.~\cite{Allen:2004vz}), remain superhorizon
early at the hot expansion epoch and get reprocessed into adiabatic
perturbations, as discussed above.

Similar situation occurs in another scenario suitable for our purposes,
namely, ``genesis'' of Ref.~\cite{Creminelli:2010ba} (see also
Ref.~\cite{starting}). According to this scenario, the Universe is
initially spatially flat and nearly static, stays in this nearly
Minkowskian state for long time, then its expansion quickly speeds up and
eventually the conventional hot epoch begins. If our conformal rolling
stage ends up well before the start of rapid expansion, the evolution of
the phase perturbations is again nearly Minkowskian up until the horizon
exit.

In both scenarios the relevant range of momenta is wide, provided
that $f_{0}$ is small enough (but not unrealistically small). We discuss
this point in Section \ref{Subsec/Pg10/1:arxiv1205/Momentum scales}. So,
it is legitimate to approximate the evolution of the phase perturbations as
Minkowskian in the time interval\footnote{For the reason that will become
clear shortly, we drop here the superscript $(0)$ in the notation of
$\eta_*$.} $\eta_* - \epsilon<\eta <\eta_1$, where $\eta_1$ is some time
after the horizon exit, and $(\eta_* -\epsilon)$ is the time when the
radial field relaxes to the minimum of $V(|\phi|)$ and the conformal
rolling stage ends. We set $\epsilon=0$ in what follows to simplify
notations; keeping $\epsilon \neq 0$ would not change our results (recall
that the phase perturbations are frozen out well before $\eta=\eta_*$).
The field $\delta \theta ({\bf x}, \eta_*)$, determined by the dynamics at
the conformal rolling stage, serves as the initial condition for further
Minkowskian evolution from $\eta_*$ to $\eta_1$. Barring fine tuning, the
case of interest for us is\footnote{In the opposite case, the phase
perturbations do not evolve between $\eta_*$ and $\eta_1$, and we are back
to the version (i) above.}
\[
k(\eta_1 - \eta_*) \gg 1 \; .
\]
Our purpose is to study the properties of the phase perturbations at
$\eta = \eta_1$, as these properties are inherited by the adiabatic
perturbations.

\begin{figure}[tb!]
\begin{center}
\includegraphics[width=0.6\textwidth,angle=0]{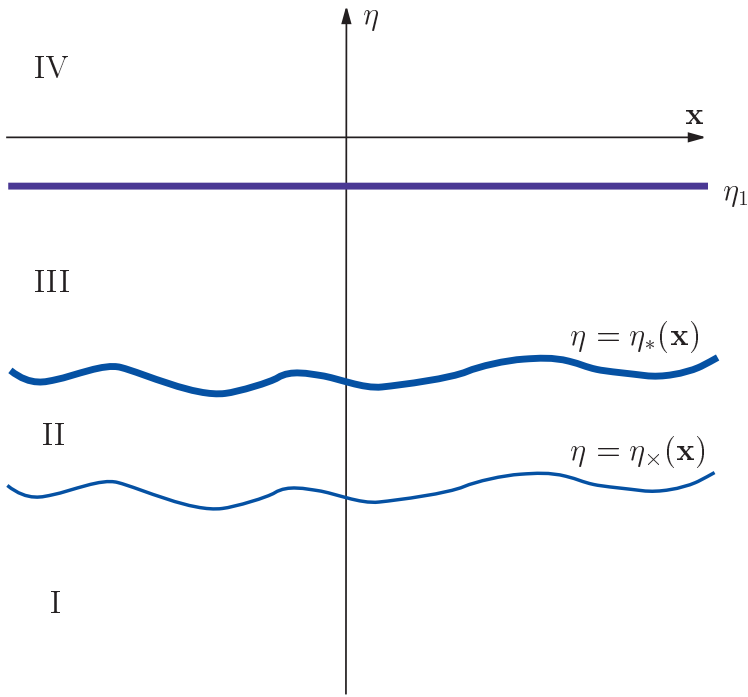}
\end{center}
\caption{Due to the perturbations of the radial field, the evolution of
phase perturbations proceeds in inhomogeneous background. Perturbations
$\delta \theta$ oscillate in time at early stage (region I), freeze out at
time $\eta=\eta_\times ({\bf x})$ and temporarily stay constant (region
II) until the end of conformal rolling that occurs at $\eta=\eta_* ({\bf
x})$. Then they evolve again, now in nearly Minkowskian regime (region
III), until the horizon exit time $\eta_1$.  Later on (region IV),
perturbations $\delta \theta$ are superhorizon and  stay constant.
\label{Figure}
 }
 \end{figure}

To the leading order in $h$, we find nothing  new: the phase perturbations
at $\eta = \eta_1$ are Gaussian and have flat power spectrum. Subleading
orders in $h$ are more interesting. A simple way to understand what is
going on is to notice that the end-of-roll time $\eta_*$, instead of being
a constant parameter, is actually a Gaussian random
field~\cite{vrscalinv}, $\eta_* ({\bf x}) = \eta_*^{(0)} + \delta \eta_*
({\bf x})$ with $\delta \eta_* \propto h$. This is due to the fact that
not only the phase $\theta$ but also the radial field $|\chi|$ acquire
perturbations at the conformal rolling stage; after freeze out,
perturbations $\delta |\chi|$ can be interpreted as perturbations $\delta
\eta_* ( {\bf x})$. The effect of the perturbations $\delta \eta_*$ on the
phase perturbations $\delta \theta$ is twofold. First, the perturbations
$\delta \eta_*$ modify the dynamics of  $\delta \theta$ at the conformal
rolling stage. This property is common to both cases (i) and (ii), and we
make use of the results of Ref.~\cite{Libanov:2010nk}. The new point is
that the resulting field $\delta \theta ({\bf x}, \eta_*({\bf x}))$ serves
as the initial condition for the Minkowskian evolution. Second, this
initial condition is now imposed at the non-trivial hypersurface $\eta =
\eta_* ({\bf x})$. This is illustrated in Fig.~\ref{Figure}.
 The net result is that the perturbation $\delta \theta ({\bf x})$  at the
 time $\eta_1$ is a combination of two Gaussian random fields originating
from vacuum fluctuations of the phase $\theta$ and radial field $|\chi|$,
respectively (better to say, from vacuum fluctuations of imaginary and
real parts of $\chi$, with our convention of real background $\chi_c$).
This leads to several potentially observable effects.

At the level of the two-point correlation function of the phase
perturbation $\delta \theta ({\bf x}, \eta_1)$, and hence of the adiabatic
perturbation $\zeta$, we have found two effects. The first one is negative
scalar tilt
\be
n_s - 1 = - \frac{3h^2}{4\pi^2} \; .
\label{jan9-1}
\ee
We note in passing that this is not a particularly strong result, as
small scalar tilt in our scenario may also originate from weak violation
of conformal invariance at the conformal rolling stage~\cite{Osipov}
and/or not exactly Minkowskian evolution of $\delta \theta$ at the
intermediate stage, cf. Appendix A. The second effect is the statistical
anisotropy: the power spectrum has the form
\be
{\cal P}_{\zeta} ({\bf k})= {\cal P}_\zeta^{(0)} (k) \left[1 +
Q(\hat{\bf k})\right] \; ,
\label{jan18-2}
 \ee
 where ${\cal P}_\zeta^{(0)}$ is independent of the direction of momentum
 (nearly flat spectrum with small tilt), $\hat{\bf k} = {\bf k}/k$ is the
unit vector along the momentum and $ Q(\hat{\bf k})$ is itself a random
field, which depends on the direction of ${\bf k}$ only. Unlike the
statistical anisotropy discussed in the inflationary
context~\cite{Peloso,aniso,soda,Shtanov}, and also in the version (i) of
the conformal rolling scenario~\cite{Libanov:2010nk}, the function $
Q(\hat{\bf k})$ contains all even angular harmonics, starting from
quadrupole. We give here the expression for $Q(\hat{\bf k})$ which
accounts for the quadrupole component only (see
Section~\ref{statanisotropy} for the results valid for all multipoles)
\be
Q(\hat{k}) =  {\cal Q} \cdot w_{ij} \left(\hat{k}_i \hat{k}_j -
\frac{1}{3} \delta_{ij}\right) \; ,
\label{jan18-1}
\ee
where $w_{ij}$ is a general symmetric traceless tensor normalized to
unity, $ w_{ij} w_{ij} = 1$, and the variance of the quadrupole component
(in the sense of an ensemble of universes) is
\be
\label{jan18-11}
\langle {\cal Q}^2 \rangle = \frac{225 h^2}{32\pi^2}
\ee
Of course, the precise values of the multipoles of  $ Q(\hat{\bf k})$
in our patch of the Universe are undetermined because of the cosmic
variance.

Due to the interaction with the perturbations $\delta \eta_*$, the
resulting phase perturbations $\delta \theta ({\bf x}, \eta_1)$ and their
descendant perturbations $\zeta$ are non-Gaussian (we leave aside here the
non-Gaussianity that may be generated at the epoch of conversion of the
phase perturbations into adiabatic ones; our scenario is not special in
this respect). Their three-point correlation function vanishes identically
due to the discrete symmetry $\theta \to - \theta$ (cf.
Ref.~\cite{Libanov:2010nk}), while the four-point correlation function has
a peculiar form
\begin{align}
\langle \zeta ({\bf k}) &  \zeta (\tilde{\bf k}) \zeta ({\bf k}^\prime)
\zeta (\tilde{\bf k}^\prime) \rangle
= \frac{  {\cal P}_\zeta^{(0)} (k)}{4\pi k^3} \frac{{\cal P}_\zeta^{(0)}
(k^\prime)}{4\pi k^{\prime \, 3}} \delta({\bf k}+\tilde{\bf k})
\delta({\bf k}^\prime+\tilde{\bf k}^\prime)
\cdot \left[ 1 + F_{NG} (\hat{\bf k}, \hat{\bf k}^\prime) \right]
\nonumber \\
& ~~~~~~~~~~~~~~~~~~~~~~~~~~~~~~~~~~~~~~~~~~~~~~ ~~~~~~~~~ + ({\bf k}
\leftrightarrow {\bf k}^\prime)  + (\tilde{\bf k} \leftrightarrow {\bf
k}^\prime) \; .
\label{jan9-2}
\end{align}
The leading term in \eqref{jan9-2} (unity in square brackets) is the
Gaussian part, while the non-Gaussianity is encoded in $F_{NG}=O(h^2)$.
 Note that the structure of the non-Gaussian part is fairly similar to
that of the disconnected four-point function. Note also that $F_{NG}$
depends on the angle between ${\bf k}$ and ${\bf k}^\prime$ only. For
reasons we discuss in Section~\ref{non-Gaussianity}, the notion of
non-Gaussianity is appropriate if the angle between ${\bf k}^\prime$ and
${\bf k}$ is small, i.e., $|\hat{\bf k} - \hat{\bf k}^\prime| \ll 1$. In
this regime, the leading behaviour of $F_{NG}$ is
 \[
 F_{NG} = \frac{3h^2}{\pi^2} \log \frac{\mbox{const}}{|\hat{\bf k} -
 \hat{\bf k}^\prime|} \; ,
 \]
where constant in the argument of logarithm cannot be reliably
calculated because of the cosmic variance. The logarithmic behavior does
not hold for arbitrarily small $|\hat{\bf k} - \hat{\bf k}^\prime|$: the
function $F_{NG} (\hat{\bf k} - \hat{\bf k}^\prime)$ flattens out most
likely at $|\hat{\bf k} - \hat{\bf k}^\prime| \sim [k(\eta_1 -
\eta_*)]^{-1/2}$, and certainly at $|\hat{\bf k} - \hat{\bf k}^\prime|
\sim [k(\eta_1 - \eta_*)]^{-1}$. So, the parameter $(\eta_1 - \eta_*)$ is
detectable in principle (but, probably, not in practice).

It is tempting to speculate that the negative scalar tilt $n_s-1 \simeq -
0.04$, favoured by the data~\cite{Larson:2010gs}, has its origin in the
dynamics we discuss in this paper. If so, our only free parameter $h$ is
determined from \eqref{jan9-1}, $h^2 \simeq 0.5$, while the small
amplitude of the adiabatic perturbations is to be attributed to the
mechanism that reprocesses the phase perturbations into adiabatic ones. In
that case the statistical anisotropy is  roughly of order 1, which is
probably inconsistent with the data. On the other hand, if one attributes
the small observed amplitude of primordial scalar perturbations,
$\sqrt{{\cal P}_\zeta} \simeq 5\cdot 10^{-5}$~\cite{Komatsu:2008hk},
entirely to the smallness of $h$, i.e., identifies ${\cal P}_{\delta
\theta}$ with ${\cal P}_\zeta$, then $h^2 \sim 10^{-7}$, and the
statistical anisotropy is at the level  $Q \sim 10^{-3}$, while the
non-Gaussianity is probably unobservable. This gives an idea of the range
of predictions of our model.

This paper is organized as follows.
We begin in Section~\ref{Subsec/Pg10/1:arxiv1205/Momentum scales} with
discussing the range of momenta of modes under study. To make the
presentation self-contained, we review in
Sections~\ref{Subsec/Pg10/1:arxiv1205/Radial perturbations} --
\ref{order-v2} the properties of the radial and phase perturbations at the
conformal rolling stage~\cite{Libanov:2010nk}. Phase perturbations at the
end of intermediate, (almost) Minkowskian stage are studied in
Section~\ref{intermed} to the first non-trivial order in $h$. This is
sufficient for evaluating the statistical anisotropy and non-Gaussianity
in Sections~\ref{statanisotropy} and \ref{non-Gaussianity}, respectively.
The calculation of the tilt \eqref{jan9-1} requires the analysis of
order-$h^2$ corrections, so we postpone it to Section~\ref{tilt}. We
discuss in Appendix~A the properties of a massless scalar field in the
contracting Universe filled with super-stiff matter. We present in
Appendices~B~--~D technical details of the calculations performed in
Sections~\ref{intermed} and \ref{statanisotropy}.

\section{Conformal rolling}
\label{rolling}

Let us review the main properties of our scalar field at the stage when it
rolls down its  potential. At this stage, the theory is described by the
action
\[
S = S_{G+M} + S_\phi \; ,
\]
where $S_{G+M}$ is the action for gravity and some matter that
dominates the evolution of the Universe, and
\be
S_\phi = \int d^4x \sqrt{-g} \left[ g^{\mu\nu}\d_\mu\phi^* \d_\nu\phi
+ \frac{R}{6} \phi^* \phi - V(\phi) \right] \;
\nonumber
\ee
is the action for the scalar field we are going to discuss. Here the
scalar potential is given by \eqref{jan12-1}. We assume that the field
$\phi$ is a spectator which does not affect the cosmological evolution;
for this reason, mixing between this field and gravitational degrees of
freedom is negligible. The background metric is assumed to be given by
\eqref{jan12-2}. One introduces the field $\chi = a\phi$, and obtains its
action in conformal coordinates in the Minkowskian form,
\be
S [\chi] = \int~d^3x~d\eta~ \left[ \eta^{\mu \nu} \d_\mu \chi^* \d_\nu
\chi + h^2 |\chi|^4 \right] \; . \nonumber
\ee
The homogeneous background solution $\chi_c (\eta)$ to the field
equation  is given by \eqref{jan5-1}. Recall that we have chosen $\chi_c$
real without loss of generality.

\subsection{Momentum scales}
\label{Subsec/Pg10/1:arxiv1205/Momentum scales}

Before discussing field perturbations in detail, let us consider momentum
scales for which our scenario, outlined in Section~\ref{intro}, is valid.
According to this scenario, conformal rolling stage ends up when the
radial field $|\phi |$ becomes of order $f_{0}$. This occurs at time $\eta
_{f}$ such that
\[
\frac{1}{a(\eta _{f})h(\eta _{f}-\eta _{*}^{(0)})}\sim f_{0}\;.
\]
Hence, the shortest waves obeying (\ref{jan12-4}) have present momenta
\[
\frac{k_{\mathrm{max}}}{a_{0}}\sim hf_{0}\cdot \frac{a(\eta
_{f})}{a_{h}}\cdot \frac{a_{h}}{a_{0}}\;,
\]
where $a_{0}$ and $a_{h}$ are the present value of the scale factor and
its value at the beginning of the hot stage, respectively.
On the other hand, we assume
that the relevant modes are subhorizon right after $\eta _{f}$,
\begin{equation}
\frac{k}{a(\eta _{f})}>H(\eta _{f})\;.
\label{Eq/Pg10/1:arxiv1205}
\end{equation}
We recall our requirement (\ref{Eq/Pg3/1:arxiv1205}) and find that the
longest waves obeying (\ref{Eq/Pg10/1:arxiv1205}) satisfy
\begin{equation}
\frac{k}{a_{0}}>\frac{k_{\mathrm{min}}}{a_{0}}\sim
\frac{hf_{0}^{2}}{M_{\mathrm{PL}}}\cdot \frac{a(\eta
_{f})}{a_{h}}\cdot\frac{a_{h}}{a_{0}}\;.
\label{Eq/Pg10/2:arxiv1205}
\end{equation}
We see that the relevant range of momenta is
\[
\frac{f_{0}}{M_{\mathrm{PL}}}\cdot k_{\mathrm{max}}<k<k_{\mathrm{max}}\;.
\]
It is wide enough, provided that the energy scale $f_{0}$ is sufficiently
low. As an example, for $k_{\mathrm{max}}/k_{\mathrm{min}}\sim
(10\mathrm{kpc})^{-1}/(10\mathrm{Gpc})^{-1}$ we need
$f_{0}<10^{-6}M_{\mathrm{PL}}$.

If our mechanism is supposed to work at contracting stage in the bouncing
Universe scanario with the hot epoch starting
immediately after bounce, the
inequality (\ref{Eq/Pg10/2:arxiv1205}) implies much stronger bound on
$f_{0}$. Indeed, $a(\eta _{f})/a_{h}>1$ in this scenario, while
$a_{h}/a_{0} \gtrsim T_{0}/T_{h}>T_{0}/M_{\mathrm{PL}}$. We require that
$k_{\mathrm{min}}/a_{0}$ is lower than the present Hubble scale $H_{0}$
and obtain
\begin{equation}
\sqrt{h}\cdot f_{0} <M_{\mathrm{PL}}\left(\frac{H_{0}}{T_{0}}
\right)^{1/2}\sim 10^{-15} M_{\mathrm{PL}}\;.
\label{Eq/Pg11/1:arxiv1205}
\end{equation}
Even for $h\sim 10^{-4}$ this implies $f_{0}<10^{6} \mathrm{GeV}$.
Interestingly, fully consistent with this scenario is the scale
$f_{0}\sim\mathrm{TeV}$.

On the contrary, $a(\eta _{f})/a_{h}$ can be large in the ``genesis''
scenario~\cite{Creminelli:2010ba}. Therefore, no bound similar to
(\ref{Eq/Pg11/1:arxiv1205}) can be established in that case.

\subsection{Radial perturbations}
\label{Subsec/Pg10/1:arxiv1205/Radial perturbations}

Let us consider perturbations about this
background. To the leading order in $h$, perturbations $\delta \chi_1 =
\sqrt{2} \delta (\mbox{Re}\, \chi)$ and $\delta \chi_2 = \sqrt{2}
\mbox{Im}\, \chi$ decouple from each other. At early times, when the
relation \eqref{jan12-3} is satisfied, the fields $\delta \chi_1$ and
$\delta \chi_2$ are free and Minkowskian. The vacuum is well defined and
we assume, as usual, that this vacuum is the initial state. The
normalization factor $\sqrt{2}$ is chosen in such a way that the real
fields $\delta \chi_1$ and $\delta \chi_2$ are canonically normalized at
early times. At late times, when the opposite inequality \eqref{jan12-4}
holds, the perturbations no longer oscillate.

We begin with the radial perturbations $\delta \chi_1$. They obey the
linearized field equation, in momentum representation,
\be
(\delta \chi_1)^{\prime \prime} +p^2~ \delta \chi_1 - 6 h^2\chi_c^2
\delta \chi_1 \equiv (\delta \chi_1)^{\prime \prime} + p^2~ \delta \chi_1
- \frac{6}{(\eta_*^{(0)} - \eta)^2} \delta \chi_1  = 0 \; ,
\label{jan12-10}
\ee
where prime denotes the derivative with respect to conformal time. We
denote the conformal momentum of the radial perturbation by ${\bf p}$ and
reserve the notation ${\bf k}$ for the conformal momentum of the phase
perturbation. The properly normalized solution to Eq.~\eqref{jan12-10} is
\[
\delta \chi_1 = \frac{1}{4\pi} \sqrt{\frac{\eta_*^{(0)} - \eta}{2}}
H_{5/2}^{(1)} \left[p (\eta_*^{(0)} - \eta) \right] \cdot \hat{B}_{\bf p}
+ h.c.\; ,
\]
where $\hat{B}_{\bf p}$, $\hat{B}_{\bf p}^\dagger$ are annihilation and
creation operators obeying the standard commutational relation
$[\hat{B}_{\bf p}, \hat{B}_{{\bf p}^\prime}^\dagger]= \delta({\bf p}-{\bf
p}^\prime)$,  $H_{5/2}^{(1)}$ is the Hankel function, and here and in what
follows in this Section we omit irrelevant phase factors. At late times
the solution approaches the asymptotics
\[
\delta \chi_1 = \frac{3}{4\pi^{3/2}} \frac{1}{p^{5/2} (\eta_*^{(0)} -
\eta)^2} \cdot \hat{B}_{\bf p} + h.c.\; .
\]
The interpretation of the behaviour $\delta \chi_1 \propto
(\eta_*^{(0)} - \eta)^{-2}$ is that the end-of-roll parameter $\eta_*$
becomes a random field. Indeed, with perturbations included, the radial
field $\mbox{Re}\, \chi= \chi_c + \delta \chi_1/\sqrt{2}$ can be written
at late times as follows,
\be
\mbox{Re}\, \chi = \frac{1}{h[\eta_* ({\bf x}) - \eta]} \; ,
\label{jan12-12}
\ee
where
\be
\eta_* ({\bf x}) = \eta_*^{(0)} + \delta \eta_* ({\bf x})
\nonumber
\ee
and the linearization in $\delta \eta_*$ is understood. The field
$\delta \eta_*$ is constant in time and is given by
\be
\delta \eta_* ({\bf x}) =
\frac{3h}{4\sqrt{2}\pi^{3/2}}\int~\frac{d^3p}{p^{5/2}} \l \mbox{e}^{i{\bf
px}} \cdot \hat{B}_{\bf p} + h.c. \r \; . \nonumber
\ee
Note that this field has red power spectrum,
\be
{\cal P}_{\delta \eta_*} = \frac{9h^2}{8\pi^2} \frac{1}{p^2} \; .
\label{jan18-30}
\ee
Clearly, the overall spatially homogeneous shift of the end-of-roll
time is irrelevant, as it can be absorbed into redefinition of the bare
parameter $\eta_*^{(0)}$. What is important is the gradient of
$\eta_*({\bf x})$, as well as higher derivatives. It is convenient to
introduce the notation
\[
v_i = - \d_i \eta_* ({\bf x}) \; .
\]
It reflects the fact that to the first order in the gradient expansion
of $\eta_*({\bf x})$ (i.e., neglecting the second derivatives of $\eta_*
  ({\bf x})$) and to the linear order in ${\bf v}$, the hypersurfaces of
constant $\mbox{Re}\, \chi$, i.e., hypersurfaces $\eta_* ({\bf x}) - \eta
=\mbox{const}$, are boosted with the velocity ${\bf v}$ with respect to
the cosmic frame: these are hypersurfaces $\eta + {\bf vx} =
\mbox{const}$. The random field ${\bf v}({\bf x})$ has flat power
spectrum, while higher derivatives of $\eta_* ({\bf x})$ have blue
spectra.

\subsection{Phase perturbations: order $v$}
\label{Subsec/Pg12/1:arxiv1205/Phase perturbations: order $v$}

Let us now turn to the perturbations $\delta \chi_2$ of the imaginary
part, and account for their interaction with radial perturbations. As we
will see in what follows, relevant perturbations $\delta \eta_*$ have
wavelengths much longer than the wavelengths of the phase perturbations,
\[
p \ll k \; ,
\]
where, as before, ${\bf p}$ and ${\bf k}$ are conformal momenta of
radial and phase perturbations, respectively. Because of this separation
of scales, it is legitimate to use the expression \eqref{jan12-12}, valid
in the late-time regime $p(\eta_* - \eta) \ll 1$, when considering the
dynamics of $\delta \chi_2$, and treat the field \eqref{jan12-12} as the
 background. It is worth noting, however, that the expression
\eqref{jan12-12} is valid to the linear order in $\delta \eta_*$ only;
furthermore, there are corrections to \eqref{jan12-12} of order $\d_i\d_j
\eta_* ({\bf x})/(\eta_* - \eta)$. Therefore, the results of this Section
are valid to order $v$ (or, equivalently, to the subleading order in $h$).
We present the expressions valid to order $v^2$ in Section~\ref{order-v2}.

With this qualification, the linearized field equation for $\delta \chi_2$
reads
\be
(\delta \chi_2)^{\prime \prime} - \d_i \d_i \; \delta \chi_2 - 2h^2
(\mbox{Re}\, \chi)^2 \cdot \delta \chi_2 \equiv (\delta \chi_2)^{\prime
\prime}
- \d_i \d_i \; \delta \chi_2 - \frac{2}{[\eta_*({\bf x} ) -  \eta]^2}
\delta \chi_2  = 0 \; .
\label{jan12-20}
\ee
At early times, when $k(\eta_* - \eta) \gg 1$, we get back to the
Minkowskian massless equation, and the solutions are spatial Fourier modes
that oscillate in time. Hence, the solution to Eq.~\eqref{jan12-20} has
the following form,
\[
\delta \chi_2({\bf x}, \eta) = \int~\frac{d^3k}{(2\pi)^{3/2}
\sqrt{2k}}~\left( \delta \chi_2^{(-)}({\bf k}, {\bf x}, \eta) \hat{A}_{\bf
k} + h.c.\right)\; ,
\]
where $\delta \chi_2^{(-)}({\bf k}, {\bf x}, \eta)$ tends to
$\mbox{e}^{i {\bf k x} - ik\eta}$ as $\eta \to -\infty$ and $\hat{A}_{\bf
k}$,  $\hat{A}_{\bf k}^\dagger$ is another set of annihilation and
creation operators. It is straightforward to see that to the linear order
in $h$ and modulo corrections proportional to $\d_i \d_j \eta_*({\bf x})$,
the solution with this initial condition is
\be
\delta \chi_2^{(-)} ({\bf k}, {\bf x}, \eta)= - \mbox{e}^{i {\bf
k}{\bf x} - ik\eta_* ({\bf x}) - i{\bf kv}(\eta_* - \eta)} \cdot
\sqrt{\frac{\pi}{2}q [ \eta_*({\bf x}) - \eta]}~ H^{(1)}_{3/2} [q
(\eta_*({\bf x}) - \eta)] \; ,
\label{jan13-5}
\ee
where $q = k + {\bf kv}$. This is basically the Lorentz boost of the
solution that one would find for $\eta_* = \mbox{const}$.

At small $\eta_* ({\bf x})- \eta$, one has $\delta \chi_2 \propto
[\eta_*({\bf x}) - \eta]^{-1}$, i.e., the same behaviour as in
\eqref{jan12-12}. So, the phase perturbation freezes out:
\be
\delta \theta ({\bf x}, \eta ) =\frac{\delta \chi_2({\bf
x},\eta)}{\mbox{Re} \, \chi ({\bf x}, \eta)} =
\int~\frac{d^3k}{\sqrt{k}}\frac{h}{4\pi^{3/2} (k+ {\bf kv})}~
\mbox{e}^{i{\bf kx} - i k\eta_*({\bf x})} \hat{A}_{\bf k} \left[1 +
O\left(\frac{\d_i \d_j \delta \eta_*}{k} \right) \right] + h.c.\; ,
\label{jan13-1}
\ee
where we again omit an irrelevant constant phase factor. Note that for
$\eta_*$ constant in space (and hence ${\bf v}=0$), i.e., to the leading
order in $h$, the phase perturbations are Gaussian random field with flat
power spectrum
\eqref{jan5-2}. The interaction with the radial perturbations makes the
situation less trivial.

The expression \eqref{jan13-1} serves as the initial condition for the
evolution of the phase perturbations at the subsequent, nearly Minkowskian
stage. We indicated in \eqref{jan13-1} that there is a correction of order
$\d_i \d_j \eta_*/k$ (the factor $k^{-1}$ is clear on dimensional
grounds). The latter correction has been calculated in
Ref.~\cite{Libanov:2010nk}; it will be irrelevant in what follows.

\subsection{Phase perturbations: order $v^2$}
\label{order-v2}

To calculate the tilt in Section~\ref{tilt}, we will need the expression
for $\delta \theta$ valid to order $v^2$, but still to the first order in
the gradient expansion of $\delta \eta_*({\bf x})$ (i.e., corrections of
order  $\d_i \d_j \eta_*/k$ are still neglected). To this end, one
observes~\cite{Libanov:2010nk} that to this order, the function
\eqref{jan12-12} is no longer a solution to the field equation. One has
instead
\be
\mbox{Re}\, \chi = \frac{1}{\gamma h[\eta_* ({\bf x}) - \eta]} \; ,
\nonumber
\ee
where $\gamma = (1-v^2)^{-1/2}$. Again using the analogy with the
Lorentz boost, one obtains, instead of \eqref{jan13-5},
\be
\delta \chi_2^{(-)} ({\bf k}, {\bf x}, \eta)= \mbox{e}^{ i
q_{||}\gamma(x_{||} + v\eta)  + i {\bf q}^T {\bf x}^T - iq\gamma \eta_*
(0)} \cdot \sqrt{\frac{\pi}{2}\gamma q [ \eta_*({\bf x}) - \eta]}~
H^{(1)}_{3/2} [\gamma q (\eta_*({\bf x}) - \eta)] \; , \nonumber
\ee
where the indices $||$ and $T$ refer to components parallel and normal
to ${\bf v}$, respectively, the boosted momenta are
\[
q_{||} = \gamma(k_{||} + kv)\; , \;\;\;\; {\bf q}^T = {\bf k}^T \; ,
\;\;\;\; q= \gamma(k + k_{||}v) \; ,
\]
and, consistently neglecting the second derivatives of $\delta \eta_*
({\bf x})$, we have used $\eta_* ({\bf x}) = \eta_* (0) - {\bf vx}$. In
the limit $q(\eta_*({\bf x}) - \eta) \to 0$ one obtains the late-time
expression for the phase, which can be written in a form, surprisingly
similar to \eqref{jan13-1}, namely
\be
\delta \theta ({\bf x, \eta}) =
\int~\frac{d^3k}{\sqrt{k}}\frac{h}{4\pi^{3/2} \gamma (k+ {\bf kv})}~
\mbox{e}^{i{\bf kx} - i k\eta_*({\bf x})} \hat{A}_{\bf k} \left[1 +
O\left(\frac{\d_i \d_j \delta \eta_*}{k} \right) \right] + h.c.\; .
\label{jan20-5}
\ee
The only difference with \eqref{jan13-1} is the factor $\gamma^{-1} =
(1-v^2)^{1/2}$ in the integrand.

\section{Evolution at intermediate stage: order $v$}
\label{intermed}

As outlined in Section~\ref{intro}, our scenario involves the evolution of
the phase perturbations $\delta \theta$ from the hypersurface
$\eta=\eta_*({\bf x})$ to the hypersurface $\eta = \eta_1 = \mbox{const}$.
At this intermediate stage, the radial field stays at the minimum of the
scalar potential, while the phase field is minimally coupled to gravity,
and  evolves in the sub-horizon regime. At time $\eta_1$, the phase
perturbations become super-horizon and freeze out again. The evolution of
the phase must be nearly Minkowskian at this stage, otherwise its power
spectrum would be grossly modified, see also Appendix~A. So, the quantity
of interest is $\delta \theta ({\bf x}, \eta_1)$, and it has to be
evaluated by solving the Minkowskian equation
\be
  \Box \delta \theta \equiv (\delta \theta)^{\prime \prime} - \d_i \d_i
  \delta \theta = 0 \; .
  \label{jan14-2}
\ee
The initial condition $\delta \theta ({\bf x}, \eta_* ({\bf x}))$ at
the hypersurface $\eta = \eta_{*}({\bf x})$ is determined by the dynamics
at the conformal rolling stage. In this Section we perform the calculation
to  the linear order in $v$, so the explicit expression is given by
\eqref{jan13-1}. The second initial condition is that the perturbation
$\delta \theta$ is frozen out by the end of the conformal rolling stage,
so that
\be
\d_N \delta \theta = 0 \;\;\;\;\; \mbox{at} \;\; \eta= \eta_* ({\bf x}) \;
,
\label{jan14-1}
  \ee
  where $\d_N$ denotes the normal derivative to the hypersurface
  $\eta= \eta_* ({\bf x})$. As pointed out in Section~\ref{intro}, the
case of interest is $k(\eta_1 - \eta_*) \gg 1$, so the evolution is long.

\subsection{Warm up}
\label{warmup}

It is instructive to begin with the unrealistic case
\[
\eta_* ({\bf x}) = \eta_* (0) - {\bf vx} \;
\]
with constant ${\bf v}$. This means that the Cauchy hypersurface  is
flat and boosted with respect to the cosmic frame. Let us consider the
solution to Eq.~\eqref{jan14-2} obeying the initial condition (cf.
\eqref{jan13-1})
\[
\delta \theta_{\bf k} ({\bf x}, \eta_*({\bf x})) = \mbox{e}^{i{\bf kx}
- i k \eta_* ({\bf x})} \; , \;\;\;\; \d_N \delta \theta_{\bf k} = 0 \; .
\]
By going to the boosted reference frame back and forth, one finds that
the solution, to the first order in $v$ (and hence in $h$), can be written
as follows,
\[
\delta \theta_{\bf k} ({\bf x}, \eta) = \mbox{e}^{i({\bf k} + k{\bf
v})[{\bf x} + {\bf v}(\eta - \eta_* (0))] - ik \eta_* (0)} \cos [(k+{\bf
kv})(\eta + {\bf vx} -  \eta_* (0))] \; .
\]
Equivalently,
\be
\delta \theta_{\bf k} ({\bf x}, \eta) = \frac{1}{2} \left[
\mbox{e}^{i({\bf k} + 2k{\bf v}){\bf x} + i(k+2{\bf kv})\eta -2i (k+{\bf
kv}) \eta_* (0)} + \mbox{e}^{i{\bf kx} - i k\eta} \right]\; .
\label{nov10-2}
\ee
The first lesson is that the solution  is the sum of waves traveling along
${\bf k}$ and (almost) in the opposite direction; we will see in what
follows that this situation is generic. Furthermore, for large enough
$(\eta - \eta_*)$ the two terms in \eqref{nov10-2} have very different
phases at given ${\bf x}$, so their interference is negligible when
integrated over ${\bf k}$ with any smooth function. The second lesson is
that the wave moving along ${\bf k}$ has momentum ${\bf k}$, while the
momentum of the wave moving in the opposite direction is $({\bf k} +
2k{\bf v})$. We interpret this as the Doppler shift. Indeed, let us go to
the reference frame $(\tau, {\bf y})$ that moves with velocity ${\bf v}$
with respect to the cosmic frame, i.e.,
\[
{\bf x} = {\bf y} - {\bf v}\tau \; , \;\;\;\;\;\; \eta = \tau - {\bf
vy} \;
\]
(recall that we work to the first order in $v$). The Cauchy
hypersurface $\eta = \eta_* ({\bf x})$ corresponds to $\tau = \eta_*(0)=
\mbox{const}$, and the mode at this hypersurface is
\[
\delta \theta_{\bf k} ({\bf y}) = \mbox{e}^{i{\bf kx} - i k \eta_*
({\bf x})} = \mbox{e}^{i({\bf k} + k{\bf v}){\bf y}} \cdot \mbox{e}^{-i(k+
{\bf kv}) \eta_* (0)} \; .
\]
The last factor here is merely a constant phase, while the first factor
describes the wave with momentum $({\bf k} + k{\bf v}) $ in the new
reference frame. In the cosmic frame, this momentum gets shifted by
$-k{\bf v}$ and $k{\bf v}$ for waves moving along ${\bf k}$ and opposite
to ${\bf k}$, respectively. Hence the result \eqref{nov10-2}. We will see
that this situation is also generic: to the first non-trivial order in
$h$, the main effect due to the intermediate stage is precisely the
Doppler shift and the lack of interference between waves coming in the
directions of ${\bf k}$ and $- {\bf k}$.

\subsection{General formula and saddle point calculation}
\label{saddlepoint}

The general solution to the Cauchy problem for Eq.~\eqref{jan14-2} with
the field and its normal derivative specified at hypersurface $\Sigma$ is
\be
\delta \theta (x) = \int_\Sigma~ d \Sigma^\mu \left\{ D^{ret} (x,y)
\d_\mu \delta \theta (y) - \left[ \frac{\d}{\d y^\mu} D^{ret} (x,y)\right]
\delta \theta (y)\right\} \; ,
\label{jan15-2}
\ee
where  $D^{ret}$ is the retarded Green's function of
Eq.~\eqref{jan14-2}, $x$ collectively denotes the coordinates $(\eta ,{\bf
x})$, and the normal to the hypersurface is directed towards future. In
our case the first term in the integrand is absent because of
\eqref{jan14-1}. We make use of the explicit expression
(valid in the case
$x^0 > y^0$ we are interested in)
\be
D^{ret}(x,y) = \frac{1}{2\pi} \delta [(x-y)^2] \; ,
\label{jan15-3}
\ee
perform the integration over the radial variable and obtain for large
$(\eta_1 - \eta_*)$ (see Appendix~B for details)
\be
\delta \theta (x) = \int~\frac{d\Omega_{\bf n}}{4\pi} \frac{1}{1-{\bf
nv}} r \d_r \delta \theta  \; ,
\label{nov10-3}
\ee
where we still use the notation $v_i = - \d_i \eta_*$. Here ${\bf n}$
is unit radius-vector, integration runs over the unit sphere parametrized
by ${\bf n}$, and $r=r({\bf n})$ is the spatial distance that light
travels from the hypersurface $\eta = \eta_*({\bf y})$ to the point $x
=(\eta _{1},{\bf x})$. It obeys the following equation:
\be
r = \eta_1 - \eta_* ({\bf x} + {\bf n}r)\; .
\label{nov10-11}
\ee
The function $\delta \theta = \delta \theta (r, {\bf n})$ in the right
hand side of \eqref{nov10-3} is the field value at the Cauchy
hypersurface,
\[
\delta \theta (r, {\bf n}) = \delta \theta({\bf y}, \eta_* ({\bf y}))
\;
\]
with
\[
{\bf y} = {\bf x} + {\bf n}r \; .
\]
The formula \eqref{nov10-3} is exact for large $r$ (for arbitrary $r$
and general Cauchy data with non-vanishing $\d_N \delta \theta$, its
generalization is Eq.~\eqref{jan15-b} in Appendix~B).

We now make use of \eqref{jan13-1} and obtain
\be
\delta \theta({\bf x}, \eta_1) = \frac{h}{4\pi^{3/2}} \int~\frac{d^3
k}{\sqrt{k}} \mbox{e}^{i {\bf kx}} A_{\bf k} \cdot I + h.c.\; ,
\label{jan17-1}
\ee
where $I$ is the integral over unit sphere,
\be
I = i \int~\frac{d\Omega_{\bf n}}{4\pi} \mbox{e}^{i \psi ({\bf n})}
\cdot r \cdot \frac{({\bf k} + k{\bf v}) {\bf n}}{(1-{\bf nv})(k +{\bf
kv})} \;
\label{nov10-20}
\ee
with
\be
\psi =  {\bf kn}r - k\eta_* ({\bf x} + {\bf n}r) = {\bf kn} \eta_1 -
({\bf kn } + k) \eta_* ({\bf x} + {\bf n}r) \; .
\label{jan16-4}
\ee
All quantities in the integrand of \eqref{nov10-20} (including ${\bf
v}$) are to be evaluated at ${\bf y} = {\bf x} + {\bf n}r$. Corrections to
 the integrand are of order $v^2$ and $\d v/k$.

The exponential factor  $\mbox{e}^{i\psi}$ in \eqref{nov10-20} is,
generally speaking, a rapidly oscillating function of ${\bf n}$, since
$\psi$ is proportional to the large parameter $kr$. Therefore, the
integral \eqref{nov10-20} can be calculated by the saddle point method,
adapted to our problem. When performing the calculation, we have to keep
in mind one point. Namely, even though we deal with soft modes in $\delta
 \eta_* ({\bf x})$ (with momenta $p\ll k$), the term  $k\eta_* ({\bf x} +
{\bf n}r)$ in $\psi$ also gives rise to a rapidly oscillating factor,
since $r$ is large. So, we cannot neglect the second derivatives $\d^2
\eta_*$ in the exponent $\psi$.

The saddle points are extrema of $\psi ({\bf n})$, where ${\bf n}$ is {\it
a unit vector}. To find them, let us formally consider ${\bf n}$ as an
arbitrary vector, and $\psi$ formally as a function of this vector. Then
the extremum on unit sphere is the point where $\d\psi/\d {\bf n}$ is
parallel to ${\bf n}$, i.e.,
\be
\frac{\d \psi}{\d {\bf n}} = \lambda kr {\bf n}
\label{nov10-10}
\ee
with yet to be determined $\lambda$ (the factor $kr$ on the right hand
side is introduced for further convenience; in fact, $\lambda kr$ is
nothing but the Lagrange multiplier). We use Eq.~\eqref{nov10-11} to find,
to the first order in $v$,
\be
\frac{\d r}{\d {\bf n}} = {\bf v} r \nonumber
\ee
and, therefore,
\be
\frac{\d \psi}{\d {\bf n}} = \left[ {\bf k} + ({\bf kn} + k) {\bf
v}\right] r \;.
\label{jan16-10}
\ee
We see that there are two saddle points, one near the unit vector
$\hat{\bf k} = {\bf k}/k$ directed along the momentum, and another near
$(-\hat{\bf k})$. These saddle points correspond to waves moving from the
Cauchy hypersurface in directions opposite to ${\bf k}$ and along ${\bf
k}$, respectively, in accord with the discussion in Section~\ref{warmup}.

The contributions of the two saddle points to the integral
\eqref{nov10-20} are calculated in Appendix~C to the first order in $v$
and $\d v$. They sum up to
\be
I = \frac{1}{2k}\left\{ \mbox{e}^{i\psi_+} \left[ 1 - \hat{\bf k}{\bf
v}^{(+\hat{\bf k})} + r (\delta_{ij} - \hat{k}_i \hat{k}_j)\d_i
v_j^{(+\hat{\bf k})}\right] + \mbox{e}^{i\psi_-} \left( 1 - \hat{\bf
k}{\bf v}^{(-\hat{\bf k})} \right) \right\} \; ,
\label{jan16-2}
\ee
where
\begin{align*}
\psi_+= \psi_+ ({\bf x}, \hat{\bf k}) &= k\eta_1 - 2k \eta_* ({\bf x} +
\hat{\bf k}r)\;, \\
\psi_- &= - k\eta_1 \; ,
\end{align*}
and superscripts $(+\hat{\bf k})$ and $(-\hat{\bf k})$ indicate that the
corresponding quantities are to be evaluated at
\begin{subequations}
\label{jan18-5}
\be
{\bf y}^{(+)}= {\bf x} + \hat{\bf k} r
\ee
and
\be
{\bf y}^{(-)}= {\bf x} - \hat{\bf k} r \; ,
\ee
\end{subequations}
respectively. The terms in \eqref{jan16-2} marked by $+$ and $-$ come from
the saddle points ${\bf n}\approx \hat{\bf k}$ and ${\bf n}\approx -
\hat{\bf k}$, respectively; they are analogs of the two terms in
\eqref{nov10-2} (the factor $(k + {\bf kv})^{-1} = k^{-1} (1- \hat{\bf
k}{\bf v})$ in the integrand in \eqref{jan13-1} was ignored in
Section~\ref{warmup}). Note that there is no symmetry between the two
contributions; technically, this is because the dependence on $\delta
\eta_*$ is absent in the phase \eqref{jan16-4} for ${\bf n} = - \hat{\bf
k}$, but present for ${\bf n} =  \hat{\bf k}$. Note also that the saddle
point value $\psi_+$ depends on ${\bf x}$ already to the linear order in
$h$, while the second saddle point value $\psi_-$ does not. This is
precisely what we observed in Section~\ref{warmup}: the momentum of
perturbation corresponding to the first contribution in \eqref{jan16-2} is
${\bf k} + \d \psi_+/\d {\bf x} = {\bf k} + 2k {\bf v}$, like in the first
term in \eqref{nov10-2}. Note finally that since we consider the case $kr
\gg 1$, it is  legitimate to neglect the correction of order $\d^2
\eta_*/k = \d v /k$, indicated in \eqref{jan13-1}, while keeping the
correction of order $r \d v$ in \eqref{jan16-2}.

One more remark is in order. Our notation ${\bf v}^{(\pm \hat{\bf k})}$
suggests that these quantities are functions of the direction of momentum
only, i.e., that they are independent of the length of the vector ${\bf
k}$. This is true, but within our approximation only. The reason is that
the horizon exit time $\eta_1$ is different for different $k$, so the
arguments ${\bf y}^{(\pm)}$ of  ${\bf v}^{(\pm \hat{\bf k})}$ depend on
$k$ through $r = \eta_1 - \eta_*$. This is irrelevant for us, since
$|\eta_1 (k) - \eta_1 (k^\prime)|$ is at most of order $1/k$ or
$1/k^\prime$ (in fact, it is even smaller, cf. Appendix~A), so the effect
we discuss is of order $\d v/k$. Also, one may worry that the phases
$\psi_{\pm}$ depend on $k$ through $\eta_1$. This is irrelevant as well,
for the following reason. When calculating the correlation functions of
the field $\delta \theta$, one neglects the interference between the
contributions due to the first and second saddle points, since the
interference term oscillates in $k$ as $\mbox{e}^{2ikr}$ and is negligible
when integrated with any smooth function of ${\bf k}$. Then the factor,
say, $\mbox{e}^{ik\eta_1}$ is merely a phase factor that can be absorbed
into the redefinition of $A_{\bf k}$. In other words, ${\bf
x}$-independent phases cancel out in the correlation functions of $\delta
\theta$, so the dependence on $k$ through $\eta_1$ does not appear. These
observations apply to all calculations in this paper, so we neglect the
dependence of $\eta_1$ on $k$ in what follows.

We conclude this Section by the discussion of the range of validity of our
saddle point calculation. It follows from \eqref{jan16-4} that the
relevant region of angular integration in \eqref{nov10-20} near each of
the saddle points is $\Delta \vartheta \sim (kr)^{-1/2}$. The saddle-point
calculation makes sense if $\eta_*({\bf x} + {\bf n}r)$ does not change
dramatically at this angular scale. Hence, by the saddle point method we
can only treat the interaction of the phase perturbations with the modes
of $\delta \eta_*$ whose momentum $p$ obeys $ pr \Delta \vartheta \lesssim
1$, i.e.,
\be
p \lesssim \sqrt{\frac{k}{r}} \; .
\label{jan20-1}
\ee
The momenta $p$ relevant for the statistical anisotropy do obey this
inequality, see Section~\ref{statanisotropy}, while the requirement
\eqref{jan20-1} restricts the angular scales at which we can reliably
study non-Gaussianity. The latter point is further discussed in the end of
Section~\ref{non-Gaussianity}.

\section{Statistical anisotropy}
\label{statanisotropy}

We see from Eqs.~\eqref{jan17-1} and \eqref{jan16-2} that the resulting
phase perturbation $\delta \theta ({\bf x}, \eta_1)$ is a combination of
two random fields, one associated with operators $A_{\bf k}$ and
$A^\dagger_{\bf k}$ and another being $\delta \eta_* ({\bf x})$. Let us
discuss the two-point product $\delta \theta ({\bf x}) \delta \theta ({\bf
x}^\prime)$ averaged over the realizations of $A_{\bf k}$ and
$A^\dagger_{\bf k}$ for {\it one realization} of $\delta \eta_*$, still to
the linear order in $h$ (in this Section we consider solely the resulting
perturbations $\delta \theta ({\bf x}, \eta_1)$ and omit the argument
$\eta_1$ in the notation). As discussed in the end of
Section~\ref{saddlepoint}, we neglect interference between terms with
$\mbox{e}^{i \psi_+}$ and  $\mbox{e}^{i \psi_-}$. Then the  two-point
function reads
\begin{align}
\langle \delta \theta ({\bf x}) \delta \theta ({\bf x}^\prime) \rangle =
\frac{h^2}{16\pi^3} & \left\{ \frac{1}{4}\int~\frac{d^3k}{k^3}
\mbox{e}^{i({\bf k } + 2k{\bf v}^{(+\hat{\bf k})}) ({\bf x} - {\bf
x}^\prime)} \cdot\left[ 1 - 2\hat{\bf k}{\bf v}^{(+\hat{\bf k})} + 2r
(\delta_{ij} - \hat{k}_i \hat{k}_j)\d_i v_j^{(+\hat{\bf k})}\right]
\right. \nonumber\\
+ & \left. \frac{1}{4}\int~\frac{d^3k}{k^3} \mbox{e}^{i{\bf k} ({\bf x}
- {\bf x}^\prime)} \cdot \left( 1 - 2\hat{\bf k}{\bf v}^{(-\hat{\bf k})}
 \right) \right\}\; ,
\label{nov11-1}
\end{align}
where we made use of the fact that, to the first order in $v$,
\[
\psi_+ ({\bf x}, \hat{\bf k})  - \psi_+ ({\bf x}^\prime, \hat{\bf k}) =
2 k{\bf v}^{(+\hat{\bf k})} ({\bf x} - {\bf x}^\prime) \; .
\]
Since we consider the long-ranged component of the field ${\bf v}$,
i.e., $p\ll k$, we neglect the terms of order $|{\bf x}- {\bf x}^\prime|
\cdot \d v$. In particular, we do not distinguish between ${\bf v}({\bf
x}^\prime + \hat{\bf k}r)$ and ${\bf v}({\bf x} + \hat{\bf k}r)$ in the
right hand side of \eqref{nov11-1}.

We now see explicitly that the actual momentum corresponding to the first
term in \eqref{nov11-1} equals ${\bf k} + 2k{\bf v}$, whereas the momentum
in the second integrand equals ${\bf k}$. To obtain the standard form of
the Fourier expansion, we  change the variable to $ \tilde{\bf k} = {\bf
k} + 2k{\bf v}$ in the first integral. To the first orger in $h$, the
Jacobian of this change of variables is
\[
\left(\mbox{det}~ \frac{\d \tilde{k}_i}{\d k_j} \right)^{-1} =1 - 2
\hat{\bf k} {\bf v}^{(+\hat {\bf k})} - 2k \frac{\d  v_i^{(+\hat {\bf
k})}}{\d k_i} = 1 - 2 \hat{\bf k} {\bf v}^{(+\hat {\bf k})} - 2 \d_j
v_i^{(+\hat {\bf k})} \cdot r (\delta_{ij} - \hat{k}_i \hat{k}_j) \; ,
\]
where we recalled that  ${\bf v}^{(+\hat {\bf k})} = {\bf v} ({\bf x} +
\hat{\bf k} r)$. It is worth noting that the last term here cancels out
the last term in square brackets in \eqref{nov11-1}. So, omitting tilde
over $\tilde{\bf k}$, we obtain that for given realization of ${\bf
v}({\bf x})$, the power spectrum, with the correction of the first order
in $h$, has the following form:
\be
{\cal P}_{\delta \theta}({\bf k}) = {\cal P}_0 \left[1 + \hat{k}_i
\left(v_i^{ (+\hat{\bf k})} -v_i^{ (-\hat{\bf k})} \right) \right] \equiv
{\cal P}_0 \left[1 + Q(\hat{\bf k}) \right]\; ,
\label{nov19-10-1}
\ee
where
\[
{\cal P}_0 = \frac{h^2}{8\pi^2}
\]
is the power spectrum to the leading order in $h$ (it is twice smaller
than the power spectrum at conformal rolling stage after freeze-out of the
phase perturbations; this is because the contributions of the two saddle
points do not sum up coherently at $\eta=\eta_1$). Note that the
non-trivial term in \eqref{nov19-10-1} depends on the {\it  direction} of
momentum ${\bf k}$. Note also that the power spectrum  \eqref{nov19-10-1}
is symmetric under ${\bf k} \to - {\bf k}$, so the two-point function
\eqref{nov11-1} is invariant under ${\bf x} \leftrightarrow {\bf
x}^\prime$, as it should. Low angular harmonics of ${\bf v}^{(\pm \hat{\bf
k})}$, viewed as a function on unit sphere in momentum space, take certain
values in our patch of the Universe. Hence, they induce statistical
anisotropy; in particular, the lowest multipole of the expression in the
right hand side of \eqref{nov19-10-1} (quadrupole) gives rise to the power
spectrum of the form \eqref{jan18-2}, \eqref{jan18-1}.

In more detail, the right hand side of  \eqref{nov19-10-1} contains all
even
multipoles,
\be
Q(\hat{\bf k}) = \sum_{lm} q_{lm} Y_{lm}(\hat{\bf k}) \; ,
\label{jan18-10}
\ee
where $Y_{lm}$ are spherical harmonics. Making use of the definition
${\bf v}^{(\pm \hat{\bf k})} = {\bf v}({\bf y}^{(\pm)})$, where ${\bf
y}^{(\pm)}$ are given in \eqref{jan18-5}, we find for $l \neq 0$ that the
multipole coefficients are given by
\begin{equation}
\label{coef}
q_{lm}=-i \int d^3p \delta \eta_{*} ({\bf{p}}) \int d
\Omega_{\hat{\bf k} } Y^{*}_{lm} (\hat{\bf k})\cdot {\bf p}\hat{\bf k}
\left( e^{ir{\bf p}\hat{\bf k}}- e^{-ir{\bf p}\hat{\bf k}} \right) \; ,
\end{equation}
 where we omitted an irrelevant $\hat{\bf k}$-independent phase.
It is worth noting that for low multipoles, the
relevant range of integration over ${\bf p}$ is roughly $p \sim r^{-1}$:
at larger $p$ the integrand rapidly oscillates, while at smaller $p$ the
expression in the inner integrand in \eqref{coef} decays as $p^2$ while
according to \eqref{jan18-30} the amplitude of $\delta \eta_* ({\bf p})$
behaves as $\sqrt{{\cal P}_{\delta \eta_*}} \propto p^{-1}$. At large $l$,
the relevant momenta are of order $p \sim l r^{-1}$. Thus, our
approximation $p \ll (k/r)^{1/2}$ is justified at least for low
multipoles.

The calculation of the variance of $q_{lm}$ is performed in much the same
way as the calculation of the CMB anisotropy multipoles, see, e.g.,
Ref.~\cite{book2}. This is done in Appendix D with the result
\be
\langle q_{lm} q_{l^\prime m^\prime}^{*} \rangle = \frac{3h^2}{\pi}
\frac{1}{(l-1)(l+2)}
\delta_{l l^\prime} \delta_{m m^\prime} \; ,
 \;\;\;\;\; ~\mbox{even}~~~l \neq 0 \; .
\label{jan18-61}
\ee
Note that we use different normalization here and in \eqref{jan18-1}.
To establish the correspondence, we calculate
the angular integral of the variance of the quadrupole term in
\eqref{jan18-1}:
\[
\int~d\Omega_{\hat{\bf k}} \langle \left[{\cal Q}
\cdot w_{ij} \left(\hat{k}_i \hat{k}_j - \frac{1}{3} \delta_{ij} \right)
\right]^2 \rangle = \frac{8\pi}{15} \langle {\cal Q}^2 \rangle \;
,
\]
while the  same integral of the quadrupole term in \eqref{jan18-10} is
given by
\[
\int~d\Omega_{\hat{\bf k}} \langle \;\vert \sum_{m=-2}^2 q_{2m} Y_{2m}
(\hat{\bf k})\vert^2 \rangle =  \sum_{m=-2}^2 \langle |q_{2m}|^2 \rangle
\; .
\]
Hence the extra factor $75/8\pi$ in \eqref{jan18-11} as compared to
\eqref{jan18-61}.

\section{Non-Gaussianity}
\label{non-Gaussianity}

The statistical anisotropy is an appropriate notion for describing the
effect due to the variation of ${\bf v}^{(\pm \hat{\bf k})}$ over {\it
large} angular scales in momentum space. On the other hand, the effect of
fluctuations of  ${\bf v}^{(\pm \hat{\bf k})}$ at {\it small} angular
scales is naturally interpreted, we believe, in terms of non-Gaussianity.
Indeed, in the latter case it makes sense to treat ${\bf v}^{(\pm \hat{\bf
 k})}$ as genuine random field and perform averaging over its
realizations, having in mind multiplicity of patches in the $\hat{\bf
k}$-sphere.

It is worth noting that even though we are going to consider  ${\bf
v}^{(\pm \hat{\bf k})}$ at small angular scales $\Delta \vartheta$ in
momentum space, the relevant momenta ${\bf p}$ of the field $\delta
\eta_*$ are still small, $p \sim (r \Delta \vartheta)^{-1}$. So, our
approximation $p \ll (k/r)^{1/2}$ is still valid, provided that $\Delta
\vartheta$ is not very small, see the discussion in the end of this
Section.

Let us consider higher order correlation functions of $\delta \theta ({\bf
x})$ (we again omit the argument $\eta_1$ in this Section). Since this
field has the general structure \eqref{jan17-1}, where $I$ does not
contain the operators $ A_{\bf k}$, $ A_{\bf k}^\dagger$, the three-point
function vanishes identically. For calculating the non-Gaussian part of
the four-point function, the expression \eqref{jan16-2}, valid to the
first order in $v$, is sufficient. Proceeding in the same way as in the
beginning of Section~\ref{statanisotropy}, we obtain
\begin{align}
\langle \delta \theta ({\bf k}) & \delta \theta (\tilde{\bf k}) \delta
\theta ({\bf k}^\prime)   \delta \theta (\tilde{\bf k}^\prime) \rangle
=   \frac{1}{4\pi k^3} \frac{1}{4\pi k^{\prime \, 3}} {\cal P}_0^2 \,
\delta({\bf k}+\tilde{\bf k}) \delta({\bf k}^\prime+\tilde{\bf k}^\prime)
\nonumber \\
& ~~~~~~\times \left[ 1 + G_{NG} (\hat{\bf k},\hat{\bf k}^\prime) +
G_{NG} (-\hat{\bf k},\hat{\bf k}^\prime)  + G_{NG} (\hat{\bf k},-\hat{\bf
k}^\prime) + G_{NG} (-\hat{\bf k},-\hat{\bf k}^\prime) \right] \nonumber
\\
& ~~~~~~~~~~~~~~~~~~~~~~~~~~~~~~~~~~~~~~~~~~~~ ~~~~~~~~~~~~~~~~~ +
({\bf k} \leftrightarrow {\bf k}^\prime)  + (\tilde{\bf k} \leftrightarrow
{\bf k}^\prime) \; ,
\label{jan19-5}
\end{align}
where the non-Gaussianity is encoded in
\be
 G_{NG} (\hat{\bf k},\hat{\bf k}^\prime)
= \langle  \hat{k}_i \left( v_i^{(+\hat{\bf k})} -v_i^{
(-\hat{\bf k})} \right) \cdot
 \hat{k}^\prime_l
\left(v_l^{(+\hat{\bf k}^\prime)} -v_l^{ (-\hat{\bf k}^\prime)} \right)
\rangle \; .
\label{jan19-1}
\ee
Fluctuations of ${\bf v}^{(\pm \hat{\bf k})}$ at small angular scales in
momentum space show up when ${\bf k}$ and ${\bf k}^\prime$ are either
nearly parallel, or nearly antiparallel, the latter case being related to
the former by the interchange ${\bf k} \leftrightarrow \tilde{\bf k}$. So,
it suffices to consider nearly parallel ${\bf k}$ and ${\bf k}^\prime$,
i.e.,
\[
|\hat{\bf k} - \hat{\bf k}^\prime| \ll 1 \; .
\]
Since the power spectrum of the random field ${\bf v}({\bf x})$ is flat,
the leading term is logarithmic in $|\hat{\bf k} - \hat{\bf
k}^\prime|$.

The expression in \eqref{jan19-1} involves the combination
\[
{\bf v}^{(+\hat{\bf k})} - {\bf v}^{ (-\hat{\bf k})} = {\bf v} ({\bf x}
+\hat{\bf k} r) - {\bf v} ({\bf x} - \hat{\bf k} r) \; .
\]
Therefore, the integral over momenta of the field ${\bf v}$ is cut off
in the infrared at $p \sim r^{-1}$. We cannot quantitatively treat the
modes of these momenta anyway, since they are plagued by cosmic variance.
 So, we consider modes with $p > r^{-1}$, recall the expression
\eqref{jan18-30} for the power spectrum of $\delta \eta_*$ and write
\[
 G_{NG} = 2 \langle \hat{k}_i v_i^{(+\hat{\bf k})} \cdot
 \hat{k}^\prime_j v_j^{ (+\hat{\bf k}^\prime)} \rangle_{p\gtrsim r^{-1}} =
2 \cdot \frac{9h^2}{8\pi^2} \int_{p\gtrsim r^{-1}}~\frac{d^3 p}{4\pi p^5}
\hat{k}_i  \hat{k}^\prime_j p_i p_j ~\mbox{e}^{i{\bf p }(\hat{\bf k} -
\hat{\bf k}^\prime)r} \; .
 \]
The angular integral here is straightforwardly evaluated. We make use
of the fact that $\hat{\bf k}(\hat{\bf k} - \hat{\bf k}^\prime) =
O((\hat{\bf k} - \hat{\bf k}^\prime)^2)$ and obtain
\[
 G_{NG} = - \frac{9h^2}{4\pi^2} \int_{x\gtrsim |\hat{\bf k} - \hat{\bf
 k}^\prime|} ~ \frac{dx}{x^2} \left(\frac{\sin x}{x} \right)^{\prime} \; ,
\]
where $x = rp|\hat{\bf k} - \hat{\bf k}^\prime|$. This is a logarithmic
integral, and in the leading logarithmic approximation we immediately get
\[
G_{NG} = \frac{3h^2}{4\pi^2} \log \frac{\mbox{const}}{|\hat{\bf k} -
\hat{\bf k}^\prime|} \; .
\]
The constant here is of order 1; it cannot be reliably calculated,
since the contribution of the region $p \sim r^{-1}$ is undetermined
because of the cosmic variance. Finally, we notice that the right hand
side of \eqref{jan19-1} is symmetric under ${\bf k} \to -{\bf k}$, so the
four terms in \eqref{jan19-5} give equal contributions. Thus, the
four-point function at $|\hat{\bf k} - \hat{\bf k}^\prime| \ll 1$ is
\begin{align}
\langle \delta \theta ({\bf k}) & \delta \theta (\tilde{\bf k}) \delta
\theta ({\bf k}^\prime) \delta \theta (\tilde{\bf k}^\prime) \rangle
= \frac{1}{4\pi k^3} \frac{1}{4\pi k^{\prime \,3}} {\cal P}_0^2 \,
\delta({\bf k}+\tilde{\bf k}) \delta({\bf k}^\prime+\tilde{\bf k}^\prime)
\cdot \left[ 1 + F_{NG} (\hat{\bf k}-\hat{\bf k}^\prime) \right] \nonumber \\
 & ~~~~~~~~~~~~~~~~~~~~~~~~~~~~~~~~~~~~~~~~~~~~ ~~~~~~~~~~~~~~~~~ +
 ({\bf k} \leftrightarrow {\bf k}^\prime)  + (\tilde{\bf k}
 \leftrightarrow {\bf k}^\prime) \; ,
\nonumber
\end{align}
where
\be
F_{NG} (\hat{\bf k}-\hat{\bf k}^\prime) = \frac{3h^2}{\pi^2} \log
\frac{\mbox{const}}{|\hat{\bf k} - \hat{\bf k}^\prime|} \; .
\label{jan19-6}
\ee
We conclude this Section by noting that our analysis is valid provided
that we can treat the range  $x \equiv rp|\hat{\bf k} - \hat{\bf
k}^\prime|\sim 1$ within our approximation $p \ll (k/r)^{-1/2}$, see
\eqref{jan20-1}. So, the logarithmic behaviour \eqref{jan19-6} persists
until $|\hat{\bf k} - \hat{\bf k}^\prime| \gtrsim (kr)^{-1/2}$. At even
smaller angles between  $\hat{\bf k}$ and $\hat{\bf k}^\prime$, the
function $F_{NG}  (\hat{\bf k}-\hat{\bf k}^\prime) $ most likely flattens
out. The logarithmic behaviour is definitely absent for $|\hat{\bf k} -
\hat{\bf k}^\prime| \lesssim (kr)^{-1}$, since momenta $p$ higher than $k$
 do not contribute to the effect. These observations suggest that the
value of $r$ is potentially measurable.

\section{Tilt}
\label{tilt}

Once the interactions of the phase field with the radial one are not
neglected, the power spectrum of the phase perturbations obtains a small
tilt. The reason is that for larger $k$, there are more modes of $\delta
\eta_*$ with $p < k$ which affect the properties of the phase
perturbations. We will see that the effect is logarithmic because of the
flat spectrum of ${\bf v}$.

To this end, let us come back to the two-point correlation function
$\langle \delta \theta ({\bf x})  \delta \theta ({\bf x}^\prime) \rangle$.
Even though the integral \eqref{nov10-3} is again saturated near ${\bf n}
= \pm \hat{\bf k}$, the saddle point calculation like that performed in
Section~\ref{saddlepoint} is no longer appropriate, since we are going to
consider all modes of $\delta \eta_*$ of momenta $p \ll k$ and not
necessarily very large wavelength modes obeying \eqref{jan20-1}. The
problem is not notoriously difficult, nevertheless, since we are
interested in logarithmically enhanced effect. Imagine that one calculates
$\langle \delta \theta ({\bf x})  \delta \theta ({\bf x}^\prime) \rangle$
by expanding in $\delta \eta_*$ the complete expression \eqref{nov10-3},
with $\delta \theta $ in the integrand given by \eqref{jan20-5}. In
principle, large logarithms could come from the expectation values
$\langle \delta \eta_* \cdot \d_i \d_j \delta \eta_*\rangle$ and $\langle
 v_i \cdot v_j\rangle$. We reiterate, however, that the overall time shift
is irrelevant for our problem, so the terms of the former type do not
appear explicitly (for the same reason, there are no terms involving
 correlation functions of $\delta \eta_*$ with itself and with ${\bf v}$,
which would yield power law corrections). Thus, it is legitimate to ignore
the correction of order $\d_i \d_j \delta \eta_*$ in \eqref{jan20-5}.
Moreover, we can formally consider the velocity ${\bf v}$ in
\eqref{jan20-5} as a constant which is independent of spatial coordinates.
So, we effectively deal with the Lorentz-boosted hypersurface $\eta_* =
\eta_*({\bf y}_\pm) - {\bf v}({\bf y} - {\bf y}_\pm)$, where ${\bf y}_\pm
= {\bf x} \pm \hat{\bf k}r$. The qualification here is that the velocity
is to be evaluated at ${\bf y} = {\bf y}_\pm$, and that ${\bf v}({\bf
y}_\pm)$ is a non-linear function of $\delta \eta_*$, since, according to
\eqref {nov10-11}, ${\bf y}_\pm$ depend on $\delta \eta_*$ through $r$.
Another qualification is that when calculating the power spectrum ${\cal
P}_{\delta \theta}$ at momentum $k$, we have to impose a restriction $p<
 k$ on the momentum $p$ of modes of the field $\delta \eta_*$.

Since we treat the velocity ${\bf v}$ as constant in space, we can obtain
the solution to the Cauchy problem explicitly, in a way similar to that of
Section~\ref{warmup}. However, we need the solution to the second order in
$v$. The initial condition for the Minkowskian evolution is thus given by
\eqref{jan20-5}. Let us define the Lorentz-boosted coordinates ${\bf z}$
and $\tau$:
\[
z_{||} = \gamma (x_{||} + v \eta) \;, \;\;  \;\;\;\;\;\;\; {\bf z}_T =
{\bf x}_T  \;, \;\;  \;\;\;\;\;\;\; \tau = \gamma (\eta + v x_{||}) \; ,
\]
where $||$ and $T$ refer to components parallel and normal to velocity.
Then the initial data are specified at the hypersurface $\tau = \tau_{\pm}
= \mbox{const}$, and $\d_\tau \delta \theta = 0$ at this hypersurface. We
re-express ${\bf x}$ and $\eta$ in terms of ${\bf z}$ and $\tau$ and
insert them into  \eqref{jan20-5}. Omitting the overall phase factor
independent of ${\bf z}$ that cancels out in the two-point function, we
write the initial conditions as
\be
\delta \theta ({\bf z}, \tau_\pm) \propto \int~ \frac{1}{\gamma (k
+{\bf kv})} \mbox{e}^{i{\bf qz}} A_{\bf k} \frac{d^3k}{\sqrt{2k}} + h.c.\;
, \;\;\;\;\; \d_\tau \delta \theta =0 \; ,
\ee
where
\[
q_{||} = \gamma (k_{||} + kv) \; , \;\;\;\;\;\; {\bf q}_T = {\bf k}_T
\; .
\]
The solution to the massless field equation in Minkowski space with
this initial condition is
\be
\delta \theta \propto \int~ \frac{1}{\gamma (k +{\bf kv})}
\mbox{e}^{i{\bf qz}} \cos \left[ q(\tau - \tau_\pm)) \right] A_{\bf k}
\frac{d^3k}{\sqrt{2k}} + h.c.\; ,
\ee
where $q = \gamma (k + k_{||}v)$. This solution again describes two
waves propagating in opposite directions, which do not interfere at $\eta
= \eta_1$. Let us consider the two waves separately.

At  time $\eta_1$, we have for the first wave, moving in direction
opposite to ${\bf k}$,
\be
\mbox{e}^{i{\bf qz} + iq \tau} = \mbox{e}^{i\gamma^2 (k_{||} + 2kv +
k_{||} v^2) x_{||} + i {\bf k}_T {\bf x}_T + i\varphi} \; ,
\ee
where $\varphi$ is a phase, irrelevant for the two-point function of
$\delta \theta$. So, the actual momentum is
\be
\tilde{k}_{||} = \gamma^2 (k_{||} + 2kv + k_{||} v^2) \; ,
\;\;\;\;\;\;\;\; \tilde{\bf k}_T = {\bf k}_T \;. \nonumber
\ee
Note that to order $v^2$ we have $\tilde{k}_{||} = \gamma_{2v} (k_{||}
+ 2kv)$, where $\gamma_{2v} = (1 -4v^2)^{-1/2}$ is the Lorentz-factor for
velocity $2v$. Hence, $\tilde{\bf k}$ again
differs from momentum ${\bf k}$ by
the Lorentz-boost with velocity $2{\bf v}$. We recall that $A_{\bf k}
{d^3k}/{\sqrt{2k}}$ is Lorentz-invariant, and obtain for the contribution
of the first wave at point ${\bf x}$ (again omitting the phase factor,
irrelevant for the two-point function)
\be
\delta \theta ({\bf x}) \propto \int~\frac{1}{\gamma (k +{\bf kv})}
\mbox{e}^{i\tilde{\bf k} {\bf x}} A_{\tilde{\bf k}}
\frac{d^3\tilde{k}}{\sqrt{2{\tilde{k}}}} + h.c.\; , \nonumber
\ee
where $ k_{||} = \gamma_{2v} (\tilde{k}_{||} - 2\tilde{k}v)$, $k =
\gamma_{2v}(\tilde{k} - 2 \tilde{k}_{||} v)$. Expanding in ${\bf v}$ to
the second order, we obtain the following form of the first contribution
to the power spectrum
\be
{\cal P}_{\delta \theta} (\tilde{k})\propto 1 +
2\left(\frac{\tilde{k}_i}{\tilde{k}} \langle v_i \rangle +
\frac{\tilde{k}_i \tilde{k}_j}{\tilde{k}^2}  \langle v_i v_j\rangle -
\frac{1}{2} \langle v^2 \rangle \right) + \frac{\tilde{k}_i
\tilde{k}_j}{\tilde{k}^2} \langle v_i v_j\rangle \;. \nonumber
\ee
Here the term in parentheses comes from $\langle \delta \theta^{(0)}
\delta\theta^{(1)} \rangle$, where $\delta \theta^{(0)}$ is the zeroth
order phase perturbation and  $\delta \theta^{(1)}$ the correction (that
includes linear and quadratic terms in ${\bf v}$), while the last term in
the right hand side is due to the correlator $\langle \delta \theta^{(1)}
\delta \theta^{(1)} \rangle$. We see that the explicitly quadratic terms
 cancel out and find (at this point we can set $\tilde{\bf k}={\bf k}$)
\be
{\cal P}_{\delta \theta} \propto 1 + 2\hat{k}_i \langle v_i \rangle \;
. \nonumber
\ee
We now recall that the velocity is to be evaluated at the point ${\bf
y}_+ = {\bf x} + \hat{\bf k}  r$, where $r = \eta_1 - \eta_*^{(0)} -
\delta \eta_* ({\bf x}  + \hat{\bf k} r) $, so that
\be
v_i ({\bf y}_+) = v_i  [{\bf x} + \hat{\bf k} (\eta_1 - \eta_*^{(0)})]
- \d_j v_i \cdot \hat{ k}_j \delta \eta_* \;. \nonumber
\ee
The expectation value of the first term on the right hand side vanishes,
while the second term gives
\be
\langle v_i ({\bf y_{+}})\rangle =  - \hat{k}_j \int \frac{d^3 p}{4\pi
p^3} p_i p_j {\cal P}_{\delta \eta_*} (p) = - \hat{k}_i\cdot \frac{3
h^2}{8\pi^2} \log (kr) \; , \nonumber
\ee
where we recalled that the relevant range of momenta is $r^{-1} \ll p
\ll k$. Thus, the contribution due to the first wave has the form
\be
{\cal P}_{\delta \theta} \propto 1 - \frac{3 h^2}{4\pi^2} \log (kr)\;
.
\label{dec23-3}
\ee

Let us now consider the second wave that moves along ${\bf k}$. We have at
time $\eta_1$
\be
\mbox{e}^{i{\bf py} - ip\tau} = \mbox{e}^{i{\bf kx} + i\varphi} \; ,
\ee
so the actual momentum is equal to ${\bf k}$. Hence, the contribution
of this wave is
\be
\delta \theta ({\bf x})\propto \int~\frac{1}{\gamma (k +{\bf kv})}
\mbox{e}^{i{\bf k z}} A_{\bf k} \frac{d^3k}{\sqrt{2k}} + h.c.\; .
\nonumber
\ee
Proceeding as before, we obtain the contribution of this wave to the
power spectrum,
\be
{\cal P}_{\delta \theta} \propto 1 - 2\hat{k}_i \langle v_i \rangle \;
, \nonumber
\ee
where the velocity is to be evaluated at the point ${\bf y}_{-} = {\bf
x} - \hat{\bf k} r$ with $r = \eta_1 - \eta_*^{(0)} - \delta \eta_* ({\bf
x}  - \hat{\bf k} r) $. The resulting contribution again has the form
\eqref{dec23-3}, so we conclude that the shape of the entire power
spectrum is given by \eqref{dec23-3}.

The result  \eqref{dec23-3} shows that the power spectrum of $\delta
\theta$, and hence of the adiabatic perturbations, is tilted. If this is
the only reason for the tilt, the scalar spectral index in our model is
equal to $n_s = 1 - \frac{3 h^2}{4\pi^2}$. As pointed out in
Section~\ref{intro}, however, there may be other sources for the tilt, so
we cannot insist on attributing the entire scalar tilt to the effect
discussed in this Section.

To end up this Section, we sketch an alternative way of calculating the
correction to the power spectrum ${\cal P}_{\delta \theta}$. One makes use
of the exact formula \eqref{nov10-3} with $\delta \theta $ in the
integrand given by \eqref{jan20-5} and evaluated at ${\bf y} = {\bf x} +
{\bf n}r$, where $r$ obeys Eq.~\eqref{nov10-11}. The dependence of the
integrand in \eqref{nov10-3} on the integration variable ${\bf n}$ is
fairly non-trivial, since the vector ${\bf n}$ enters the argument of
$\eta_*$ both explicitly and through $r({\bf n})$. We know, however, that
the integral \eqref{nov10-3} is saturated in regions near the two points
on unit sphere, ${\bf n} = \hat{\bf k}$ and ${\bf n} = -\hat{\bf k}$.
Consider the first region for definiteness. The idea is to write
\begin{align*}
\eta_*[{\bf x} + {\bf n}r({\bf n})] &= \eta_*[{\bf x} + \hat{\bf k}r
(\hat{\bf k})] + \left\{ \delta \eta_*[{\bf x} + {\bf n}r({\bf n})] -
\delta \eta_*[{\bf x} + \hat{\bf k}r (\hat{\bf k})] \right\}\;, \\
r({\bf n}) &= r(\hat{\bf k}) - \left\{ \delta \eta_*[{\bf x} + {\bf
n}r({\bf n})] - \delta \eta_*[{\bf x} + \hat{\bf k}r (\hat{\bf k})]
\right\} \; ,
\end{align*}
express these functions iteratively through
\[
\delta \eta_*[{\bf x} + {\bf n}r(\hat{\bf k})] - \delta \eta_*[{\bf x}
+ \hat{\bf k}r(\hat{\bf k})] \;
\]
and systematically expand the integrand in \eqref{nov10-3} in a series
in the latter quantity, up to quadratic order. Then one has to deal with
angular integrals, in which the integration variable ${\bf n}$ enters
either in combination $\mbox{e}^{i{\bf k}{\bf n}r(\hat{\bf k})}$ or via
$\mbox{e}^{i{\bf k}{\bf n}r(\hat{\bf k})} \delta \eta_* [ {\bf x} + {\bf
n}r(\hat{\bf k})] $ (the integral with $\delta \eta_*^2$ is trivial after
ensemble averaging). The former integral is straightforwardly evaluated by
the saddle point method. To evaluate the latter integral, one writes
$\delta \eta_* [ {\bf x} + {\bf n}r(\hat{\bf k})] $ in the Fourier
representation and arrives at the angular integral with $\mbox{e}^{i({\bf
k}+ {\bf p}){\bf n}r(\hat{\bf k})} $, where ${\bf p}$ is still the
momentum of a mode of $\delta \eta_*$. The latter integral is again
evaluated by the saddle point method; the rest of the calculation is
straightforward.

We have performed the calculation of the power spectrum in this way; it is
tedious, but does yield  the result \eqref{dec23-3}.

\section*{Acknowledgements}

The authors are indebted to W.~Buchm\"uller, A.~Hebecker, D.~Levkov,
M.~Osipov, G.~Rubtsov, M.~Sazhin, S.~Sibiryakov and Ch.~Wetterich for
useful comments and discussions. We are particularly grateful to
Y.~Shtanov who pointed out an inconsistency in the first version of this
paper. This work has been supported in part by the Federal Agency for
Science and Innovations under state contract 02.740.11.0244 and by the
grant of the President of the Russian Federation NS-5525.2010.2. The work
of M.L. has been supported in part by IISN and by Belgian Science Policy
(IAP VI/11) and by Russian Foundation for Basic Research grant
 11-02-92108. The work of S.R.  has been supported in part by  the grant
of the President of the Russian Federation MK-7748.2010.2 and
MK-3344.2011.2 and by Federal Agency for Education under state contract
P520. M.L. and S.R. acknowledge the support by the Dynasty Foundation.
M.L. thanks Service de Physique Th\'{e}orique,  Universit\'{e} Libre de
Bruxelles, where part of this work has ben done, for hospitality. V.R.
thanks Institute of Theoretical Physics, Heidelberg University where part
 of this work has been done, for hospitality.

\section*{Appendix A. Scalar field perturbations in a contracting
Universe with super-stiff matter}

In this Appendix we discuss the free propagation of massless scalar field
$\Phi$ minimally coupled to gravity in contracting Universe filled with
matter whose equation of state is super-stiff, $p=w\rho$, $w \gg 1$. Since
the phase $\theta$ behaves precisely in this way after freeze out of the
radial field $|\phi|$, our discussion applies directly to the situation
studied in this paper. Our point is to show that in the limit $w \to
\infty$ the propagation is effectively Minkowskian all the way down to
$a(\eta) \to 0$.

For constant $w$, the scale factor evolves in conformal time as follows,
\[
  a = |\eta|^{\beta} \; , \;\;\;\;\;\; \eta<0 \; ,
  \]
  where
  \[
  \beta = \frac{2}{1+3w} \; .
  \]
  In terms of the field $\sigma = a \Phi$, the field equation reads
  \be
  \sigma^{\prime \prime} + k^2 \sigma - \frac{a^{\prime \prime}}{a}
  \sigma = \sigma^{\prime \prime} + k^2 \sigma +
\frac{\beta(1-\beta)}{\eta^2}\sigma =0 \; .
  \label{jan11-1}
\ee
For large $w$ and hence small $\beta$, the last term in the left hand
side of Eq.~\eqref{jan11-1} is negligible before the horizon exit time,
$\eta_{ex} \sim - \sqrt{\beta}/{k}$, while there is simply no time to
evolve even in Minkowski space in the time interval $(\eta_{ex},0)$. This
is why one can make use of the Minkowskian evolution to evaluate the value
of the field $\Phi$ as $\eta \to 0$, i.e., deep in the super-horizon
regime.

To substantiate this claim, let us consider the Cauchy problem similar to
that discussed in the main text. Namely, let the initial value  $\Phi_i$
be specified at $\eta=\eta_* = \mbox{const}$ with $|\eta_*| \gg k^{-1}$,
and another initial condition is $\Phi^\prime =0$ at $\eta=\eta_*$. Let us
compare the values of $\Phi$ obtained at $\eta=0$ by solving the
Minkowskian evolution equation $\Box \Phi = 0$ and by evolving the field
according to Eq.~\eqref{jan11-1}. The Minkowski evolution gives
$\Phi^{Mink} (\eta) = \Phi_i \cos k(\eta - \eta_*)$, so that
\[
\Phi^{Mink} (\eta \to 0) = \Phi_i \cos k\eta_* \; .
\]
The solution to Eq.~\eqref{jan11-1} with the above initial conditions
imposed at $|\eta_* | \gg k^{-1}$ is
\[
\sigma (\eta) = \Phi_i |\eta _{*}|^{\beta }\sqrt{\frac{\pi}{2} k|\eta|}
\left[u H_\nu^{(1)} (-k\eta) + u^* H_\nu^{(2)} (-k\eta) \right] \; ,
\]
where $\nu = 1/2 - \beta$,
\[
u = \frac{1}{2} \mbox{e}^{ik\eta_* + i \frac{\pi}{2}\left(1 -
\beta\right)}
\]
and $H^{(1,2)}_\nu$ are Hankel functions. The asymptotics of $\Phi =
\sigma/a$ as $\eta \to 0$ for $\beta <1/2$ is
\[
\Phi(\eta \to 0) = \Phi_i \cos\left(k\eta_* - \frac{\pi \beta}{2}
\right) \left(\frac{k|\eta_*|}{2}\right)^\beta \frac{\Gamma(1/2-\beta
)}{\Gamma(1/2)} \; .
\]
We see that the Minkowskian result indeed coincides with the exact one
in the limit $w \to \infty$, i.e., $\beta \to 0$. The main effect for
finite but large $w$ is the induced tilt in the power spectrum. The phase
$\pi \beta/2$ is irrelevant, as it cancels out in the correlation
functions.

\section*{Appendix B. Derivation of the formula~\eqref{nov10-3}}

In this Appendix we consider the Cauchy problem for Eq.~\eqref{jan14-2}
with initial data specified at the hypersurface
\be
  f(y) = \eta - \eta_* ({\bf y}) = 0 \; ,
  \label{jan15-4}
  \ee
  where  $y$ denotes a point with coordinates $y^\mu = (\eta, {\bf
  y})$. We simplify the notation and use $\theta (x)$ instead of $\delta
\theta (x)$.

Let $\tilde{\theta} (x)$ be the solution to the D'Alembert equation
\eqref{jan14-2}, such that $\tilde{\theta} (y)$ and $\d_N \tilde{\theta}
(y)$ coincide with the Cauchy data $\theta (y)$ and $\d_N \theta (y)$ at
the Cauchy hypersurface (hereafter $\d_N$ denotes the normal derivative).
Let us introduce
\[
\theta (x) = \tilde{\theta}(x) \cdot \Theta[f(x)] \; ,
\]
where $\Theta$ is a step function. Then
\[
\Box \, \theta = \d_\mu \tilde{\theta}\, \d^\mu f \cdot \delta(f) +
\d_\mu [\tilde{\theta} \,\d^\mu f \cdot \delta(f)]
\]
and, therefore,
\be
\theta (x) = \int~d^4y~\left\{D^{ret}(x,y) \, \d_\mu \theta (y)\,
\d^\mu f(y) \cdot \delta[f(y)] - \left[\frac{d}{dy^\mu} D^{ret}(x,y)
\right] \theta (y) \, \d^\mu f(y) \cdot \delta[f(y)]\right\} \; ,
\label{jan15-1}
\ee
where we omitted tilde over $\theta$ in the right hand side, since the
integration runs over the Cauchy hypersurface. The second term in the
integrand is obtained by integration by parts. The formula \eqref{jan15-1}
is nothing but the general formula \eqref{jan15-2}, and $\d_\mu \theta \,
\d^\mu f \propto \d_N \theta$.

In the case of interest, the normal derivative vanishes at the Cauchy
hypersurface, and the first term in the integrand in \eqref{jan15-1} is
absent. We make use of \eqref{jan15-3} and write
\[
\frac{d}{dy^\mu} D^{ret}(x,y) = - \frac{1}{\pi} (x_\mu - y_\mu)
\delta^\prime [(x-y)^2] \; .
\]
We use the explicit form \eqref{jan15-4} of $f(y)$, integrate over
$\eta$ in \eqref{jan15-1} and obtain for $x=(\eta_1,{\bf x})$
\be
\theta (x) = \frac{1}{\pi} \int~d^3y~ [\eta_1 - \eta_*({\bf y}) + {\bf
v}({\bf x} - {\bf y})]\,  \theta({\bf y})\, \delta^\prime \left([\eta_1 -
\eta_*({\bf y})]^2 -({\bf x} - {\bf y})^2 \right) \; ,
\label{jan15-6}
\ee
where $v_i = -\d_i \eta_* ({\bf y})$ and $\theta({\bf y}) \equiv
\theta [{\bf y}, \eta_*({\bf y})]$ is the field value at the Cauchy
hypersurface. We now introduce the integration variable ${\bf r}$ via
${\bf y} = {\bf x} + {\bf r}$, write ${\bf r} = {\bf n}r$, where ${\bf n}$
is a unit vector, and cast the integral \eqref{jan15-6} into the following
form:
\be
\theta (x) = \frac{1}{\pi} \int~d\Omega_{\bf n}\, r^2 dr~ [\eta_1 -
\eta_*({\bf x} + {\bf n}r) - {\bf nv}r] \, \theta({\bf x} + {\bf n}r) \,
\delta^\prime \left([\eta_1 - \eta_*({\bf x} + {\bf n}r)]^2 - r^2\right)
\; .
\label{jan15-7}
\ee
Here ${\bf v} = {\bf v}({\bf x} +{\bf n}r)$. Finally, we make use of
the identity
\[
\delta^\prime \left([\eta_1 - \eta_*({\bf x} + {\bf n}r)]^2 -
r^2\right) = - \frac{1}{2 \left\{r - {\bf nv}[\eta_1 - \eta_*({\bf x} +
{\bf n}r)] \right\}} \frac{\d}{\d r} \delta \left([\eta_1 - \eta_*({\bf x}
+ {\bf n}r)]^2 - r^2\right) \; ,
\]
which is obtained by evaluating the derivative over $r$ of $\delta
\left([\eta_1 - \eta_*({\bf x} + {\bf n}r)]^2 - r^2\right)$. Since $r\neq
0$ at the Cauchy hypersurface, we can integrate over $r$ in
\eqref{jan15-7} by parts. We also use the fact that
\be
\delta \left([\eta_1 - \eta_*({\bf x} + {\bf n}r)]^2 - r^2\right) =
\frac{1}{2r(1-{\bf nv})} \delta [r - r({\bf n})]\;,
\label{jan15-8}
\ee
where $r({\bf n})$ is the solution to Eq.~\eqref{nov10-11}. We get
\begin{align*}
\theta (x) = \frac{1}{\pi} \int~d\Omega_{\bf n} dr~ \frac{\d}{\d r}
&\left( \frac{r^2}{ 2 \left\{r - {\bf nv}[\eta_1 - \eta_*({\bf x} + {\bf
n}r)] \right\}} [\eta_1 - \eta_*({\bf x} + {\bf n}r) - {\bf nv}r]\,
\theta({\bf x} + {\bf n}r) \right) \\
& ~~~~~~\times \frac{1}{2r(1-{\bf nv})} \delta [r - r({\bf n})] \;.
\end{align*}
The integration over $r$ is now straightforward, and we obtain after some
algebra (note the cancellation of the terms with derivative $\d {\bf v}
({\bf x} + {\bf n}r)/\d r$)
\be
\theta (x) = \frac{1}{4\pi} \int~d\Omega_{\bf n} \left[\theta +
\frac{1}{1-{\bf nv}} r \d_r \theta \right] \; ,
\label{jan15-a}
\ee
where in the right hand side one has $\theta = \theta({\bf y},
\eta_*({\bf y}))$ with ${\bf y} = {\bf x} + {\bf n}r$. Let us emphasize
that \eqref{jan15-a} is the exact result for the Cauchy problem with $\d_N
\theta =0$. At large $r$, the second term in the integrand dominates, and
we arrive at the formula \eqref{nov10-3} used in the text.

For completeness, let us derive the general formula for the solution to
the Cauchy problem with non-vanishing $\d_N \theta$. With the Cauchy
hypersurface defined by Eq.~\eqref{jan15-4}, the derivative along the unit
normal is given by
\be
\d_N \theta = \gamma \d_\mu \theta \, \d^\mu f \; ,
\label{jan15-10}
\ee
where $\gamma = (1-v^2)^{-1/2}$. This expression can be obtained by
performing local boost
\[
d \tau = \gamma (d\eta + {\bf v}d{\bf x}) \; , \;\;\;\; \mbox{etc.}
\]
Then $\tau$ is the time coordinate along the normal, and
\[
\d_N \theta = \d_\tau \theta = \gamma (\d_\eta \theta - v_i \d_i
\theta)\; ,
\]
which is precisely \eqref{jan15-10}.  Making use of \eqref{jan15-10}
and \eqref{jan15-3} we write the first term in \eqref{jan15-1} as follows,
\[
\int~d^3y \frac{1}{2\pi} \delta \left( [\eta_1 - \eta_*({\bf y})]^2 -
({\bf x} -{\bf y})^2\right) \frac{1}{\gamma} \d_N \theta \; .
\]
 We proceed as before, again use \eqref{jan15-8} and obtain for this
 term
 \[
\frac{1}{4\pi} \int~d\Omega_{\bf n} ~ \frac{r}{\gamma (1-{\bf nv})}
\d_N \theta \; .
\]
Thus, the complete expression for the solution to the Cauchy problem is
\be
\theta (x) = \frac{1}{4\pi} \int~d\Omega_{\bf n} \left[\theta +
\frac{1}{1-{\bf nv}} r \left(\d_r \theta + \sqrt{1-v^2} \d_N \theta
\right) \right] \; .
\label{jan15-b}
\ee
The notations here are the same as in \eqref{jan15-a}.

\section*{Appendix C. Details of saddle point calculation}

\subsection*{Saddle point ${\bf n} \approx \hat{\bf k}$}

To find the saddle points of the integral \eqref{nov10-20}, we solve
Eq.~\eqref{nov10-10} with $\d \psi /\d {\bf n}$ given by \eqref{jan16-10}.
To the linear order in $h$, the first saddle point is
\[
{\bf n}_+ = \hat{\bf k} + 2[{\bf v} -  \hat{\bf k} \cdot (\hat{\bf
k}{\bf v})]
\]
with
\be
\lambda = 1+2 \,\hat{\bf k}{\bf v} \; .
\label{jan16-1}
\ee
Let us evaluate the contribution to the integral \eqref{nov10-20}
coming from the saddle point region near ${\bf n}_+$. Let $\vartheta$,
$\varphi$ be angular coordinates in the frame with the third axis along
${\bf n}_+$. Then
\[
{\bf n} = {\bf n}_+ + {\bf n}^{(1)} + {\bf n}^{(2)} \; ,
\]
where ${\bf n}^{(1)}$ and   ${\bf n}^{(2)}$ are of the first and second
order in $\vartheta$, respectively,
\begin{align*}
{\bf n}^{(1)} &= (\sin \vartheta \cos \varphi,  \sin \vartheta \sin
\varphi, 0)\;, \\
{\bf n}^{(2)} &= (0, 0, \cos \vartheta -1)\;.
\end{align*}
We have
\[
\psi({\bf n}) = \psi ({\bf n}_+) + \psi^{(2)}  \;,
\]
where
\[
\psi^{(2)} = \frac{\d \psi}{\d n_i} n_i^{(2)} + \frac{1}{2} \frac{\d^2
\psi}{\d n_i \d n_j} n^{(1)}_i n^{(1)}_j
\]
and the derivatives are evaluated at ${\bf n} = {\bf n}_+$. The first
derivative is given by Eqs.~\eqref{nov10-10} and \eqref{jan16-1}, while to
the linear order in $v$ and $\d v$ (i.e., linear order in $h$), the second
derivative is
\be
\frac{\d^2 \psi}{\d n_i \d n_j} = r (k_i v_j + k_j v_i) + r^2({\bf
k}{\bf n}_+ +k)  \d_i v_j \; . \nonumber
\ee
The angular integral is now straightforwardly  evaluated (one first
integrates over $\vartheta$ near $\vartheta=0$ with weight $\vartheta
d\vartheta$, then expands in $v$ and $\d v$ and integrates over
$\varphi$), and to the linear order in $h$ one finds
\be
i r \int~ \frac{d \Omega_{\bf n}}{4\pi} \mbox{e}^{i \psi^{(2)}} =
\frac{1}{2k} [1 -2(\hat{\bf k} {\bf v}) ] [1 + r (\delta_{ij} - \hat{k}_i
\hat{k}_{j}) \d_i v_j ] \; . \nonumber
\ee
The pre-exponential factor in \eqref{nov10-20} is to be evaluated at
${\bf n} = {\bf n}_+$. Collecting all factors, we get the contribution of
the first saddle point (to the first order in $h$):
\be
I_+ = \frac{1}{2} \mbox{e}^{i \psi ({\bf n}_+)} \frac{1 + r  \cdot
(\delta_{ij} - \hat{k}_i \hat{k}_{j}) \d_i v_j }{k + {\bf kv}}\;.
\nonumber
\ee
Note a non-trivial cancellation between ${\bf v}$-dependent terms  in
the pre-exponential factor. Finally, we recall that
\be
\psi ({\bf n}_+) = {\bf k}{\bf n}_+ \eta_1 - ({\bf k}{\bf n}_+ + k)
\eta_* ({\bf x} + {\bf n}_+r) = k \eta_1 - 2 k \eta_* ({\bf x} + \hat{\bf
k} r) \; ,
\nonumber
\ee
where we still work to the linear order in $h$. Since $\delta \eta_*$
and ${\bf v}$ are already of order $h$, their argument is merely ${\bf
y}^{(+)}={\bf x} + \hat{\bf k} r$. In this way we arrive at the first term
in \eqref{jan16-2}.

\subsection*{Second saddle point}

The second saddle point is precisely at
\[
{\bf n}_- = - \hat{\bf k}
\]
(this is exact result valid to all orders in $v$). At this saddle point
we have
\[
\psi ({\bf n}_-) = - k \eta_1 \; .
\]
The same calculation as above gives for the contribution of the second
saddle point
\[
I_- = \frac{1}{2} \mbox{e}^{i \psi ({\bf n}_-) } \frac{1}{k + {\bf
kv}}\; .
\]
So, the second term in \eqref{jan16-2} is obtained in a very
straightforward way.

\section*{Appendix D. Multipoles of statistical anisotropy.}

The field $\delta \eta_* ({\bf x})$ is an isotropic Gaussian field.
Therefore, the multipole coefficients in \eqref{jan18-10} are independent,
\[
\langle  q_{lm} q_{l^\prime m^\prime}^{*} \rangle = Q_l \delta_{l
l^\prime} \delta_{m m^\prime} \; .
\]
We make use of the expression \eqref{coef} and
calculate the sum
$
\sum_m \langle  |q_{lm}|^2 \rangle $.
Since $\langle \delta
\eta_* ({\bf p}) \delta \eta_*^* ({\bf p}^\prime) \rangle \propto \delta
({\bf p} - {\bf p}^\prime)$, this sum  has the following
 form:
\be
\sum_m \langle  |q_{lm}|^2 \rangle = \int~d^3p~ \frac{{\cal P}_{\delta
\eta_*}}{4\pi p^3} \sum_m |q_{lm} ({\bf p})|^2  \; .
\label{jan18-51}
\ee
The integrand here is independent of the direction of ${\bf p}$ and
therefore can be calculated in any reference frame. To simplify formulas,
we choose, somewhat loosely, a reference frame one step earlier, in the
inner integral in \eqref{coef}, so we calculate $q_{lm} ({\bf p})$ in a
${\bf p}$-dependent frame. This procedure is legitimate as long as one
calculates the sum in the right hand side of \eqref{jan18-51}. We choose
the spherical frame with ${\bf p}$ directed along the third axis and write
\begin{align}
q_{lm}({\bf p}) &= -i \int d \Omega ~Y^{*}_{lm}
(\vartheta,\varphi)\cdot p\cos \vartheta \cdot (\mbox{e}^{ipr\cos
\vartheta}-\mbox{e}^{-ipr\cos \vartheta })
\nonumber \\
&= -i \delta_{m0} \sqrt{(2l+1)\pi} \int \limits_{-1}^{1} dt
P_{l}(t) \cdot pt \cdot (\mbox{e}^{iprt} - \mbox{e}^{-iprt})
\; ,
\label{jan18-52}
\end{align}
where $P_l$ are the Legendre polynomials,
$\vartheta$ is the angle between the momenta ${\bf{p}}$ and
${\bf{k}}$ and $t =\cos \vartheta$.
Since the integrand in \eqref{jan18-52} is symmetric under
$t \to - t$ (this is a consequence of the symmetry of the power
spectrum ${\cal P}_{\delta \theta} ({\bf k})$ under
${\bf k} \to - {\bf k}$, see \eqref{nov19-10-1}), odd multipoles
vanish. In what follows we consider even $l \neq 0$.

The standard way of calculating the
integral \eqref{jan18-52}
is to make use of the expansion of the oscillating exponent in
Legendre polynomials,
\begin{equation}
\mbox{e}^{iprt}=\sum^{\infty}_{l'=0}
(2l'+1)i^{l'}j_{l'}(pr)P_{l'}(t),
\nonumber
\end{equation}
where $j_{l}$ are spherical Bessel functions.
We make use of the
normalization of the Legendre polynomials,
\begin{equation}
\int_{-1}^{1} dt P_{l}(t)P_{l'}(t)=\frac{2}{2l+1}\delta_{ll'} \; ,
\nonumber
\end{equation}
and recurrence relation
\be
tP_{l'}(t)=\frac{l'P_{l'-1}(t)+(l'+1)P_{l'+1}(t)}{2l'+1}\; . \\
\nonumber
\ee
Then the integral \eqref{jan18-52} is straightforwardly
evaluated,
\[
 q_{lm} ({\bf p})= 2 \delta_{m0} \sqrt{\frac{4\pi}{2l+1}}\, i^{l}
\, p \left[(l+1)j_{l+1}(y)- lj_{l-1} (y) \right] \; ,
\]
where
\[
y = rp \; .
\]
We now insert this result into \eqref{jan18-51}, recall that the power
spectrum of $\delta \eta_*$ is given by \eqref{jan18-30} and get
\be
\sum_{m}\langle|q_{lm}|^2\rangle=\frac{18h^2}{\pi (2l+1)}
\int_{0}^{\infty} \frac{dy}{y}
\left[(l+1)j_{l+1}(y) - lj_{l-1}(y)\right]^2 \; .
\label{jan18-62}
\ee
Finally, we recall the relationship between the spherical and
conventional Bessel functions,
\be
j_{l} (y)=\sqrt{\frac{\pi}{2y}}J_{l+\frac{1}{2}}(y)
\nonumber
\ee
and perform integration by using
\begin{equation}
\int_{0}^{\infty} J_{\nu} (y) J_{\mu} (y) y^{-\lambda} dy =
\frac{\Gamma (\lambda) \Gamma \left(\frac{\nu +\mu-\lambda +1}{2}
\right)}{2^{\lambda} \Gamma \left(\frac{-\nu +\mu +\lambda+1}{2}\right)
\Gamma \left(\frac{\nu +\mu +\lambda +1}{2} \right) \Gamma \left(\frac{\nu
-\mu +\lambda +1}{2}\right)} \; .
\nonumber
\end{equation}
After straightforward algebra this yields
\[
\sum_{m}\langle|q_{lm}|^2\rangle=\frac{3h^2}{\pi} \frac{2l+1}{(l-1)(l+2)}\; , \ \ \
\mbox{even}~~~ l>0\;,
\]
or, equivalently, the quoted result \eqref{jan18-61}.

It is worth noting that the relevant integration region in the integral
\eqref{jan18-62} is $y \equiv pr \sim l$ (the
spherical Bessel function $j_l(y)$ is exponentially small at $y \ll l$ and
decays as $y^{-1}$ at $y \gg l$). This means that our approximation $p \ll
(k/r)^{1/2}$ is justified for $kr \gg 1$, unless $l$ is very large.


\begin{thebibliography}{99}
\bibitem{inflation}
  A.~A.~Starobinsky,
  JETP Lett.\  {\bf 30}  (1979), 682;
  [Pisma Zh.\ Eksp.\ Teor.\ Fiz.\  {\bf 30} (1979), 719];
  Phys.\ Lett.\  B {\bf 91} (1980), 99.\\
  A.~H.~Guth,
  Phys.\ Rev.\  D {\bf 23} (1981), 347.\\
A.~D.~Linde,
Phys.\ Lett.\  B {\bf 108} (1982), 389;
  Phys.\ Lett.\  B {\bf 129} (1983), 177.\\
A.~Albrecht and P.~J.~Steinhardt,
Phys.\ Rev.\ Lett.\  {\bf 48} (1982), 1220.

\bibitem{infl-perturbations}
  V.~F.~Mukhanov and G.~V.~Chibisov,
  JETP Lett.\  {\bf 33} (1981), 532;
  [Pisma Zh.\ Eksp.\ Teor.\ Fiz.\  {\bf 33} (1981), 549].\\
  S.~W.~Hawking,
  Phys.\ Lett.\  B {\bf 115}  (1982), 295.\\
  A.~A.~Starobinsky,
  Phys.\ Lett.\  B {\bf 117} (1982), 175.\\
  A.~H.~Guth and S.~Y.~Pi,
  Phys.\ Rev.\ Lett.\  {\bf 49} (1982), 1110.\\
  J.~M.~Bardeen, P.~J.~Steinhardt and M.~S.~Turner,
{Phys.\ Rev.}\  D {\bf 28} (1983), 679.


\bibitem{Linde:1996gt}
  A.~D.~Linde and V.~F.~Mukhanov,
  {Phys.\ Rev.}\  D {\bf 56} (1997), 535;
   astro-ph/9610219.\\
  K.~Enqvist and M.~S.~Sloth,
  {Nucl.\ Phys.}\  B {\bf 626} (2002), 395;
   hep-ph/0109214.\\
  D.~H.~Lyth and D.~Wands,
  {Phys.\ Lett.}\  B {\bf 524} (2002), 5;
   hep-ph/0110002.\\
  T.~Moroi and T.~Takahashi,
  {Phys.\ Lett.}\  B {\bf 522} (2001), 215;
  [{Erratum-ibid.}\  B {\bf 539} (2002), 303];
   hep-ph/0110096.

\bibitem{vrscalinv}
  V.~A.~Rubakov,
  {JCAP} {\bf 0909} (2009), 030;
   arXiv:0906.3693 [hep-th].


\bibitem{Creminelli:2010ba}
 P.~Creminelli, A.~Nicolis and E.~Trincherini,
  JCAP {\bf 1011}, 021 (2010);
  arXiv:1007.0027 [hep-th].



\bibitem{Dimopoulos:2003az}
  K.~Dimopoulos, D.~H.~Lyth, A.~Notari and A.~Riotto,
  {JHEP} {\bf 0307}, (2003), 053;
   hep-ph/0304050.

\bibitem{Dvali:2003em}
  G.~Dvali, A.~Gruzinov and M.~Zaldarriaga,
  {Phys.\ Rev.}\  D {\bf 69} (2004), 023505;
   astro-ph/0303591.\\
  L.~Kofman,
  { astro-ph/0303614}.


\bibitem{Dvali:2003ar}
  G.~Dvali, A.~Gruzinov and M.~Zaldarriaga,
  {Phys.\ Rev.}\  D {\bf 69} (2004), 083505;
   astro-ph/0305548.

\bibitem{Mukohyama:2009gg}
  S.~Mukohyama,
  {JCAP} {\bf 0906} (2009), 001;
   arXiv:0904.2190 [hep-th].


\bibitem{voloshin}
M.~B.~Voloshin and A.~D.~Dolgov,
  Sov.\ J.\ Nucl.\ Phys.\  {\bf 35} (1982) 120
  [Yad.\ Fiz.\  {\bf 35} (1982) 213].


\bibitem{Libanov:2010nk}
  M.~Libanov and V.~Rubakov,
  JCAP {\bf 1011} (2010), 045;
  arXiv:1007.4949 [hep-th].

\bibitem{YKIS}
  M.~Libanov, S.~Mironov and V.~Rubakov,
  arXiv:1012.5737 [hep-th].


\bibitem{smooth}
  J.~K.~Erickson, D.~H.~Wesley, P.~J.~Steinhardt and N.~Turok,
  Phys.\ Rev.\  D {\bf 69} (2004) 063514;
  hep-th/0312009.\\
D.~Garfinkle, W.~C.~Lim, F.~Pretorius and P.~J.~Steinhardt,
  Phys.\ Rev.\  D {\bf 78} (2008) 083537;
  arXiv:0808.0542 [hep-th].

\bibitem{ekpyro}
J.~L.~Lehners,
  Phys.\ Rept.\  {\bf 465} (2008) 223;
  arXiv:0806.1245 [astro-ph].

\bibitem{ekpyro-i}
  J.~Khoury, B.~A.~Ovrut, P.~J.~Steinhardt and N.~Turok,
  Phys.\ Rev.\  D {\bf 64} (2001) 123522;
  hep-th/0103239.\\
J.~Khoury, B.~A.~Ovrut, N.~Seiberg, P.~J.~Steinhardt and N.~Turok,
  Phys.\ Rev.\  D {\bf 65} (2002) 086007;
  hep-th/0108187.

\bibitem{ekpyro-pert-old}
D.~H.~Lyth,
  Phys.\ Lett.\  B {\bf 524} (2002) 1;
  hep-ph/0106153.\\
R.~Brandenberger and F.~Finelli,
  JHEP {\bf 0111} (2001) 056;
  hep-th/0109004.




\bibitem{starting}
P.~Creminelli, M.~A.~Luty, A.~Nicolis and L.~Senatore,
  JHEP {\bf 0612} (2006) 080;
  hep-th/0606090.

\bibitem{minus-bis}
E.~I.~Buchbinder, J.~Khoury and B.~A.~Ovrut,
  Phys.\ Rev.\  D {\bf 76} (2007) 123503;
  hep-th/0702154.

\bibitem{minus-bis2}
P.~Creminelli and L.~Senatore,
  JCAP {\bf 0711} (2007) 010;
  hep-th/0702165.

\bibitem{Allen:2004vz}
  L.~E.~Allen and D.~Wands,
  Phys.\ Rev.\  D {\bf 70} (2004) 063515;
  astro-ph/0404441.

\bibitem{Osipov}
V.~Rubakov and M.~Osipov,
  arXiv:1007.3417 [hep-th].

\bibitem{Peloso}
A.~E.~Gumrukcuoglu, C.~R.~Contaldi and M.~Peloso,
  arXiv:astro-ph/0608405;
  JCAP {\bf 0711} (2007) 005; 0707.4179 [astro-ph].


\bibitem{aniso}
L.~Ackerman, S.~M.~Carroll and M.~B.~Wise,
  Phys.\ Rev.\  D {\bf 75} (2007) 083502
  [Erratum-ibid.\  D {\bf 80} (2009) 069901];
  astro-ph/0701357.\\
A.~R.~Pullen and M.~Kamionkowski,
  Phys.\ Rev.\  D {\bf 76} (2007) 103529;
  arXiv:0709.1144 [astro-ph].


\bibitem{soda}
M.~A.~Watanabe, S.~Kanno and J.~Soda,
  Phys.\ Rev.\ Lett.\  {\bf 102} (2009) 191302;
  arXiv:0902.2833 [hep-th];
 Prog.\ Theor.\ Phys.\  {\bf 123}, 1041 (2010);
  arXiv:1003.0056 [astro-ph.CO].\\
T.~R.~Dulaney and M.~I.~Gresham,
  Phys.\ Rev.\  D {\bf 81} (2010) 103532;
  arXiv:1001.2301 [astro-ph.CO].\\
A.~E.~Gumrukcuoglu, B.~Himmetoglu and M.~Peloso,
  Phys.\ Rev.\  D {\bf 81} (2010) 063528;
  arXiv:1001.4088 [astro-ph.CO].

\bibitem{Shtanov}
G.~V.~Chibisov and Yu.~V.~Shtanov,
  Sov.\ Phys.\ JETP {\bf 69} (1989) 17
  [Zh.\ Eksp.\ Teor.\ Fiz.\  {\bf 96} (1989) 32];
  Int.\ J.\ Mod.\ Phys.\  A {\bf 5} (1990) 2625.\\
R.~V.~Buniy, A.~Berera and T.~W.~Kephart,
  Phys.\ Rev.\  D {\bf 73} (2006) 063529;
  hep-th/0511115.\\
J.~F.~Donoghue, K.~Dutta and A.~Ross,
  Phys.\ Rev.\  D {\bf 80} (2009) 023526;
  astro-ph/0703455.\\
C.~Armendariz-Picon,
  JCAP {\bf 0709} (2007) 014;
  arXiv:0705.1167 [astro-ph].\\
T.~S.~Pereira, C.~Pitrou and J.~P.~Uzan,
  JCAP {\bf 0709} (2007) 006;
  arXiv:0707.0736 [astro-ph];
C.~Pitrou, T.~S.~Pereira and J.~P.~Uzan,
  JCAP {\bf 0804} (2008) 004;
  arXiv:0801.3596 [astro-ph].\\
Y.~Shtanov and H.~Pyatkovska,
  Phys.\ Rev.\  D {\bf 80} (2009) 023521;
  arXiv:0904.1887 [gr-qc];
Y.~Shtanov,
  Annalen Phys.\  {\bf 19} (2010) 332;
  arXiv:1002.4879 [astro-ph.CO].





\bibitem{Larson:2010gs}
  D.~Larson {\it et al.},
  arXiv:1001.4635 [astro-ph.CO].

\bibitem{Komatsu:2008hk}
  E.~Komatsu {\it et al.}  [WMAP Collaboration],
  Astrophys.\ J.\ Suppl.\  {\bf 180} (2009) 330;
  arXiv:0803.0547 [astro-ph].

\bibitem{book2} D.S.~Gorbunov and V.A.~Rubakov,
{\it Introduction to the Theory of the Early Universe.
 Cosmological Perturbations and Inflationary Theory},
 World Scientific, 2011.


\end{thebibliography}
\end{document}